\documentstyle[12pt,aasms]{article}
\tightenlines
 
\begin{document}
 
Accepted to ApJ -- Dec. 22, 1998
 
\vskip 48pt
 
\title 
{New Three Micron Spectra of Young Stellar Objects with H$_2$O
Ice Bands}
 
\vskip 48pt
 
\author
{T. Y. Brooke$^ {1}$}

\vskip 8pt 
 
\affil
{Jet Propulsion Laboratory, MS 169-237, 4800 Oak Grove Dr.,
Pasadena, CA  91109  USA}

\affil
{e-mail: tyb@lassie.jpl.nasa.gov}
 
\vskip 12pt

\author
{K. Sellgren$^ {1,2}$}
 
\vskip 8pt
 
\affil
{Department of Astronomy, Ohio State University,
174 West 18th Av., Columbus, OH  43210  USA}

\affil
{e-mail: sellgren@astronomy.mps.ohio-state.edu}

\vskip 12pt

\author
{T. R. Geballe$^3$}

\vskip 8pt

\affil
{Joint Astronomy Centre, 660 N. A'ohoku Pl., Hilo, HI 96720 USA}

\affil
{e-mail: t.geballe@gemini.edu}
 
\vskip 12pt
 
\altaffiltext{1}
{Visiting astronomer at Kitt Peak National Observatory, which is
operated by AURA, Inc., under cooperative agreement with the National
Science Foundation.}

\vskip 12pt

\altaffiltext{2}
{Visiting astronomer at the United Kingdom Infrared Telescope, which
is operated by the Joint Astronomy Centre on behalf of the UK Particle
Physics and Astronomy Research Council.}

\vskip 12pt

\altaffiltext{3}
{Present address: Gemini Observatory, 670 N. A'ohoku Pl., Hilo, HI 96720}

\vfill\eject
 
\centerline
{\bf Abstract}
 
We present new groundbased 3 $\mu$m spectra of 14 young stellar
objects with H$_2$O ice absorption bands.  The broad absorption
feature at 3.47 $\mu$m was detected toward all objects and its optical
depth is correlated with the optical depth of H$_2$O ice,
strengthening an earlier finding.  The broad absorption feature at
3.25 $\mu$m was detected toward two more sources and an upper limit is
given for a third source.  The optical depths of the 3.25 $\mu$m
feature obtained to date are better correlated with the optical depth
of the refractory silicate dust than with that of H$_2$O ice.  If this
trend is confirmed, this would support our proposed identification of
the feature as the C--H stretch of aromatic hydrocarbons at low
temperature.  An absorption feature at 3.53 $\mu$m due to solid
methanol was detected for the first time toward MonR2/IRS2, as well as
toward W33A and GL 2136.  The wavelengths of the CH$_3$OH features
toward W33A, GL 2136, and NGC7538/IRS9 can be fit by CH$_3$OH-rich
ices, while the wavelength of the feature toward MonR2/IRS2 suggests
an H$_2$O-rich ice environment.  Solid methanol abundances toward GL
2136, NGC7538/IRS9, and MonR2/IRS2 are 3-5 \% relative to H$_2$O ice.
There is an additional narrow absorption feature near 3.47 $\mu$m
toward W33A.  For the object W51/IRS2, spatially resolved spectra from
2 to 4 $\mu$m indicate that the H$_2$O ice is located predominantly in
front of the eastern component and that the H$_2$O ice extinction is
much deeper than previously estimated.  For the object RNO 91, spectra
from 2 to 4 $\mu$m reveal stellar (or circumstellar) CO gas absorption
and deeper H$_2$O ice extinction than previously estimated.

\vskip 12pt

\clearpage
 
\section
{Introduction}
 
\vskip 24pt

    Several distinct spectral absorption features have been detected
in the long-wavelength wing of the interstellar 3.1 $\mu$m H$_2$O ice
absorption band seen toward molecular cloud sources.  As the wing
extends through the 3.2-3.6 $\mu$m region characteristic of C--H
stretch vibrations, some or all of these features could be due to
organic species in the solid state.  Spectra of embedded young
stellar objects in this region have so far revealed a broad 3.47
$\mu$m feature and a broad 3.25 $\mu$m feature, neither of which is
securely identified yet, and a feature at 3.53 $\mu$m due to solid
CH$_3$OH.  There is also a reported detection of a narrow feature near
3.47 $\mu$m, attributed to solid H$_2$CO, toward GL 2136 (Schutte {\it
et al.} 1996a).  The species responsible for the extended
long-wavelength wing itself is still not yet certain.

   The broad 3.47 $\mu$m feature (FWHM$\approx0.1$ $\mu$m) was first
noticed in four objects by Allamandola et al. (1992). They suggested
that the feature might be due to the C--H stretch absorption of solo
hydrogens attached to $sp^3$ bonded carbon clusters, the
``diamond''-like form of carbon.  The feature was present in every
molecular cloud source looked at by Brooke, Sellgren, \& Smith (1996),
with an optical depth that appeared to be correlated with the depth of
the H$_2$O ice band, as opposed to that of the silicates.  The feature
has also been seen toward Elias 16 (Chiar, Adamson, \& Whittet 1996),
a star behind the Taurus cloud, so the presence of the feature does
not appear to require any heating or radiation processing by an
embedded protostar.  Graham (1998) detected the feature in two sources
in the RCrA association.  Under the assumption that the feature is due
to C--H bonds, Brooke, Sellgren, \& Smith (1996) interpreted the correlation
with H$_2$O ice as indicating that both C--H bonds and H$_2$O ice form
in step on molecular cloud dust by hydrogen addition reactions.
However, they noted that other identifications of the feature are also
possible.

   The broad 3.25 $\mu$m feature was first noted by Sellgren, Smith,
\& Brooke (1994) in a spectrum of MonR2/IRS3, and later confirmed with
better spectra (Sellgren {\it et al.} 1995).  They speculated that the
feature might be due to aromatic hydrocarbons at low temperature.  Two
more detections were reported (Brooke, Sellgren, \& Smith 1996), but
there were insufficient data to determine any correlations with H$_2$O
ice or silicates.  The feature falls in a wavelength region where
there are strong H$_2$O and CH$_4$ absorption lines in the terrestrial
atmosphere.  It also falls on the steeply sloping edge of the H$_2$O
ice band.  Thus reliable detections of the feature are extremely
difficult.

   Solid methanol has now been seen toward several embedded sources in
molecular clouds.  Solid CH$_3$OH was first identified in the
interstellar medium by Grim {\it et al.} (1991).  They observed an
absorption feature at 3.53 $\mu$m toward W33A, in good agreement with
the $\nu_3$ symmetric CH$_3$ stretch mode in solid CH$_3$OH.  This
feature was subsequently detected toward two other sources:
NGC7538/IRS9 (Allamandola {\it et al.} 1992) and GL 2136 (Schutte {\it
et al.} 1996a), and has recently been detected toward another,
RAFGL7009S (Dartois {\it et al.} 1998).  Additional features which can
be attributed to solid CH$_3$OH have been detected toward GL 2136 at
8.9 and 9.7 $\mu$m (Skinner {\it et al.} 1992) and toward W33A and
RAFGL7009S at 3.85 and 3.94 $\mu$m (Geballe 1991; Allamandola {\it et
al.} 1992; Dartois {\it et al.} 1998).  Also, part of the broad 6.8
$\mu$m feature seen toward many objects is most likely due to solid
CH$_3$OH (Schutte {\it et al.} 1996b and references therein).  The
wavelengths of peak absorption have been used to infer that solid
CH$_3$OH is found in ice mixtures which are not dominated by H$_2$O
ice (Skinner {\it et al.}  1992; Allamandola {\it et al.} 1992;
Schutte {\it et al.}  1996a; Dartois {\it et al.} 1998).

   As infrared instrumentation continues to improve, it has become
possible to detect weak spectral features in an increasing number of
young stellar objects.  New 3 $\mu$m spectra of 14 objects with H$_2$O
ice bands were obtained in 1995 and 1996 with two 256 $\times$ 256
InSb array spectrometers to determine the prevalence and abundance of
organics in molecular cloud dust.  These were objects with right
ascensions $16^{\rm h} < \alpha < 21^{\rm h}$, with the exception of
MonR2/IRS2.  MonR2/IRS2 was re-observed to obtain a spectrum of better
quality than that shown in Brooke, Sellgren, \& Smith (1996), which
suffered from poor cancellation of telluric features.

\vskip 12pt

\section
{Observations}

\vskip 12pt
 
   Spectra were obtained at the Kitt Peak National Observatory 2.1-m
telescope on 12 January 1996 and 4,6 June 1996 and at the United
Kingdom Infrared Telescope (UKIRT) on 6 September 1995, 12-13 June and
16 July 1996.  All dates are UT.  Table 1 is a log of the
observations.  The spectra are shown in Figs. 1-4.

   At Kitt Peak, the infrared cryogenic spectrometer CRSP was used.
The pixels were $0.61^{\prime\prime}$ across.  The slit was
$1.6^{\prime\prime}$ wide and oriented N--S for MonR2/IRS2 and E--W
for the others.  Background subtraction for MonR2/IRS2 was done by
nodding the telescope $10^{\prime\prime}$ S along the slit.  For the
other objects, the telescope was nodded $160^{\prime\prime}$ E (W for
Serpens SVS 20).  The spectra were sampled every 1/2 of a resolution
element.  Observations of MonR2/IRS2 and Serpens SVS 20 in the 3.5
$\mu$m region were taken with the 300 lines/mm grating (Grating 1),
which provided a spectral resolution $\lambda$/$\Delta
\lambda$~$\approx$~1300 at 3.5 $\mu$m.  The spectrum of SVS20 was
smoothed to an effective resolution of $\approx800$ to improve the
signal-to-noise.  Observations of W51/IRS2 and RNO 91 were taken with
the 150 l/mm grating (Grating 3), which provided $\lambda$/$\Delta
\lambda$~$\approx$~530 in the 3.4-3.9 $\mu$m region and $\approx700$ in
the 2.2 $\mu$m region.  The spectrum of RNO 91 was smoothed to an
effective resolution of $\approx200$ in the 3.6-3.9 $\mu$m region and
$\approx400$ in the 2.2 $\mu$m region.

   At UKIRT, the facility spectrometer CGS4 was used.  The pixels were
$1.2^{\prime\prime}$ across.  A $1.2^{\prime\prime}$--wide slit was
oriented E--W.  The sources were nodded $\sim12^{\prime\prime}$ along
the slit for background subtraction.  The spectra were sampled every
1/3 of a resolution element.  Spectra in September 1995 were obtained
with the 75 l/mm grating which provided a spectral resolution of
$\approx1200$.  The rest of the spectra were obtained with the 150
l/mm grating which provided a spectral resolution of $\approx$~2400.
They were smoothed to an effective resolution of $\approx1300$.

    For those sources observed at 2.2 $\mu$m, the telescope was first
positioned to the 3.5 $\mu$m peak, since surrounding nebulosity could
cause the peak position to be a function of wavelength.

   The stars used for atmospheric correction were also used for flux
calibration. The stars were assumed to be blackbodies.  The assumed
blackbody temperatures based on spectral type are given in Table 1.
The absolute flux calibration may be uncertain by 20 \%.  Where
possible, Kitt Peak spectra were scaled to the flux level of the UKIRT
spectra, which we consider more reliable.  The scaling factors were
less than 20 \%.

   Hydrogen absorption lines in the standard stars were not removed.
This means that apparent emission features at 3.04 $\mu$m (HI
Pf$\epsilon$) and 3.30 $\mu$m (HI Pf$\delta$) in the data for Elias
29, GL 2136, SVS 20, RCrA/IRS1, and RCrA/IRS2 may be due all or in
part to corresponding absorption features in the standards.  In the
case of RNO 91, the standard star used to correct the K(2.2 $\mu$m)
and 3.7 $\mu$m regions observed at Kitt Peak was an A2V star with
strong hydrogen absorptions at Br$\gamma$ (2.17 $\mu$m) and Pf$\gamma$
(3.74 $\mu$m), so these regions were deleted.  The emission lines in
the data of W51/IRS2(E) are predominantly intrinsic to the source.

   Some data points in the strongest telluric absorption features were
not well-corrected and were dropped.  Other gaps in the spectra are
due to occasional bad pixels.  Error bars are omitted in the flux
plots for clarity, but representative error bars are shown in the
plots of optical depth that follow.

\vskip 12pt

\section
{2-4 $\mu$m Spectra}

    Two young stellar objects were observed from 2 to 4 $\mu$m in
order to obtain improved estimates of the H$_2$O ice band optical
depth.  These spectra are shown in Fig. 4 and discussed in this
section.  Our goal was not an in-depth study of the 2 $\mu$m spectra
of these objects, but some interesting aspects of the spectra are
noted.

\vskip 12pt

\subsection
{W51/IRS2}
 
\vskip 12pt

   W51/IRS2 is an infrared source that contains an embedded HII
region.  It has a 3 $\mu$m H$_2$O ice absorption band previously
estimated to have an optical depth $\tau\approx0.8$ (Soifer, Russell,
\& Merrill 1976; Joyce \& Simon 1982).  These ice band observations
were taken with aperture diameters of $17^{\prime\prime}$ and
$11^{\prime\prime}$, respectively.  The source was shown to be double
in the mid-infrared by Genzel {\it et al.}  (1982).  The eastern
source, which is brighter over the 2.2 to 20 $\mu$m range, has a 10
$\mu$m silicate absorption feature, but the western source
$\approx5^{\prime\prime}$ arcsec away has no detectable silicate
feature.  Goldader and Wynn-Williams (1994) showed that W51/IRS2(E)
has a 3 $\mu$m ice absorption band but that W51/IRS2(W) does not.  Our
spectra show that the ice band toward W51/IRS2(E) is deeper than the
previous estimate.

   The 2-4 $\mu$m spectrum of W51/IRS2(E) is shown in Fig. 4a.  The
measured fluxes for W51/IRS2(E) are within 20 \% of those of Goldader
\& Wynn-Williams (1994).  The slits were oriented E-W and centered on
W51/IRS2(E).  With this orientation, we also obtained a spectrum of a
bright region about $2^{\prime\prime}$ S of the position given for
W51/IRS2(W) by Goldader \& Wynn-Williams.  Emission lines of HI, HeI
and H$_2$ were detected toward both regions.  The detected lines are
indicated in the figure for W51/IRS2(E) (Fig. 4a).  The equivalent
width of the HI Br $\gamma$ line (corrected for absorption in the
standard star) in the region $2^{\prime\prime}$ S of IRS2(W),
W$_{\lambda}$(Br$\gamma$)$\approx540$ \AA, is very close to the
theoretical value for emission by thermal electrons at T=10,000 K in
an HII region ($\approx590$ \AA, Wynn-Williams {\it et al.}  1978),
but in W51/IRS2(E) it is lower (W$_{\lambda}$(Br$\gamma$)$\approx140$
\AA), indicating that the 2 $\mu$m continuum of W51/IRS2(E) contains
starlight, scattered light and/or dust thermal emission.  This is in
agreement with the conclusion of Goldader \& Wynn-Williams (1994) from
spectrophotometric imaging.  Line ratios are given in the caption to
Fig. 4a.

  W51/IRS2(E) exhibits a strong H$_2$O ice band.  There was no
evidence for any H$_2$O ice absorption in the western region.  The
continuum chosen for the W51/IRS2(E) ice band is shown in Figure 4a.
The other curve is a scaled blackbody to indicate the approximate
color temperature of 450 K.  The local continuum adopted for the 3.47
$\mu$m feature is also shown.  The H$_2$O ice is partially crystalline
as indicated by the sharp absorption peak at 3.1 $\mu$m and the
inflections at 2.97 and 3.20 $\mu$m.  The latter inflection precludes
an accurate test for any 3.25 $\mu$m feature using our current method.
The ice band optical depth plot (Fig.  5a) shows these crystalline ice
features clearly.  The derived optical depth is $\tau(3.1)=2.4\pm0.2$,
higher than the previous estimate cited above, presumably because the
earlier measurements were contaminated by light from W51/IRS2(W).  The
true optical depth toward the source might be even higher if scattered
light or light from unresolved sources which does not undergo the same
amount of ice extinction is present in the aperture.  This would not
be unreasonable since the source is quite distant (d=7500 pc, Goldader
\& Wynn-Williams 1994) and even a small effective aperture will
subtend an extended area at the source.

\vskip 12pt
\subsection
{RNO 91}

\vskip 12pt

    RNO 91, in the L43 dark cloud, is a young stellar object which
illuminates a reflection nebula.  It has a 3 $\mu$m H$_2$O ice
absorption band previously estimated to have an optical depth
$\tau=1.29$ from a 2.8-3.8 $\mu$m spectrum (Weintraub {\it et al.}
1994).  The K(2.2 $\mu$m) band flux is extended in two lobes
(Weintraub {\it et al.}  1994).  All of our spectra were taken at the
peak 3.5 $\mu$m position, presumably the exciting star.  The K band
spectrum (Fig. 4b) shows hot CO gas overtone absorption longward of
2.3 $\mu$m.

    To estimate the ice band depth, a 3rd order polynomial (in the
log) was drawn between 2.1-2.3 and 3.8-3.9 $\mu$m, avoiding the CO
absorption (Fig. 4b).  The derived ice band optical depth profile is
shown in Fig. 5b.  The H$_2$O ice toward this source is mostly
amorphous.  The derived optical depth is $\tau(3.1)=2.1\pm0.2$, higher
than the estimate of Weintraub {\it et al.} (1994).  Our optical depth
is probably a better estimate as the 2 $\mu$m spectral continuum was
included.

    This is the first published 2 $\mu$m spectrum of this object to
our knowledge.  Because the goal was simply to define the continuum,
the integration time was short and the signal-to-noise ratio low.
However CO gas overtone absorptions can clearly be seen.  Many young
stellar objects show strong CO overtone absorption, comparable to
late-type, main-sequence giants, though they have low total
luminosities (Casali \& Eiroa 1996; Greene \& Lada 1996).  The
combined equivalent width of the CO 2--0 and 4--2 bands at 2.29 and
2.35 $\mu$m in RNO 91, defined as in Greene \& Lada (1996), is
$\approx15$ \AA.  Two other photospheric features, the atomic Na and
Ca lines at 2.21 and 2.26 $\mu$m, are weak, with equivalent widths
$\la2$ \AA.  With these line strengths, RNO 91 falls between main
sequence dwarf and giant photospheres, as do several young stellar
objects in the $\rho$ Oph cloud (Greene \& Lada 1996).  As the
bolometric luminosity of RNO 91 is low ($\approx3.7$ L$_{\sun}$, Chen
{\it et al.} 1995), the observations are consistent with photospheric
absorption by a young star characterized by low surface gravity.
Higher resolution spectra of this object should be able to confirm
whether the CO absorption bands are in fact photospheric or due to a
circumstellar disk.
%
%
%
%

\vskip 24pt
 
\section
{Determination of Spectral Features}
 
\vskip 24pt
 
   The method of defining the local continuum in order to extract the
3.25 $\mu$m and 3.47 $\mu$m features is discussed in Sellgren, Smith,
\& Brooke (1994), Sellgren {\it et al.} (1995), and Brooke, Sellgren,
\& Smith (1996).  To summarize the standard case, the local continuum
was defined to be represented by points in the ranges 3.13-3.17
$\mu$m, 3.33-3.37 $\mu$m, and points longward of 3.61 $\mu$m.  The
continuua were estimated using low order polynomial fits to the
logarithms of the flux densities of the continuum points, with the
constraint that the best fit fall above those points held to contain
features and look ``reasonable'' to the eye.  In most cases, the best
fit was a single 3rd order polynomial in the log or the joining of two
3rd order polynomials.  In cases where there were clearly no strong
broad 3.47 $\mu$m or solid CH$_3$OH features, the continuum was
allowed to extend down to 3.58 $\mu$m to improve the baseline at the
longer wavelengths.  The adopted continuua are shown in Figs. 1-4.

    An exception to this standard case was made when the 3.53 $\mu$m
absorption feature of solid CH$_3$OH was clearly present: W33A, GL
2136, and MonR2/IRS2.  The C--H stretch bands of solid CH$_3$OH absorb
strongly between approximately 3.33 $\mu$m and 3.65 $\mu$m (Hudgins
{\it et al.} 1993).  For the sources with solid CH$_3$OH, only points
near 3.33 $\mu$m were used to define the continuum in that region.  At
the long wavelength end, our spectra lacked sufficient wavelength
coverage.  We used data from previously published spectra to help
define the continuum, after first scaling the earlier spectra to the
appropriate flux level.  The earlier spectra were from Allamandola
{\it et al.} (1992) for W33A, Schutte {\it et al.} (1996a) for GL
2136, and Smith, Sellgren \& Tokunaga (1989) for MonR2/IRS2.  The data
taken from these spectra are indicated in Figs. 1a, 1b, and 3a.  The
points used to define the continuum were from 3.65 $\mu$m to 3.70
$\mu$m.

    The resulting optical depths for sources with no detected CH$_3$OH
are shown in Figs. 6 and 7.  The 3.47 $\mu$m feature and the 3.25
$\mu$m feature (where detected) were fitted with gaussians with the
central wavelength, peak optical depth, and full-width at half-maximum
(FWHM) as free parameters.  Uncertainties in these parameters were
estimated from the standard deviations of the results for fits using
other baselines that looked reasonable.  These are larger than the
formal statistical uncertainties associated with the point-to-point
scatter.  Tables 2 and 3 summarize the results.  There is no particular
significance attached to the use of gaussians; the purpose of the
fitting is only to extract reasonable estimates of the parameters of
symmetric features.  For W33A, GL2136, and MonR2/IRS2, the fits in the
3.47 $\mu$m region consisted of the sum of a gaussian and a CH$_3$OH
ice template, as discussed below.

Although a local continuum could be drawn in the 3.25 $\mu$m region in
most cases by our technique, the cancellation of telluric features was
often insufficient to put any useful constraints on a 3.25 $\mu$m
feature.  One exception was Elias 29.  No feature was detected but a
3$\sigma$ limit of $\tau(3.25)<0.03$ was estimated (Fig. 7c).

It is important to note that the actual absorption profiles of the
species responsible for the features may extend further up into the
long wavelength wing. Since we fit only a local continuum, our
technique is sensitive only to the excess absorption at 3.47 $\mu$m or
3.25 $\mu$m below this continuum, and the derived optical depths may
be lower limits to the true contributions of the absorbers.

\vskip 12pt

\section
{Notes on Some Individual Sources}

\vskip 12pt

\noindent
{\bf W33A}-- Our estimate of the optical depth of the 3.47 $\mu$m
feature is $0.29\pm0.04$, considerably higher than the value of 0.15
estimated by Allamandola {\it et al.} (1992).  This is due to the fact
that the points we held to be continuum at the long-wavelength end
were at longer wavelengths to take into account the presence of solid
CH$_3$OH.  We continue to consider this value to be a lower limit to
the true line-of-sight value, as the shape of the 3.1 $\mu$m H$_2$O
ice band suggests that light suffering less ice extinction, possibly
scattered light near the source or unresolved sources, is filling in
the band to some extent (Brooke, Sellgren, \& Smith 1996).  The 3.25
$\mu$m region was too faint to fit due to the deep H$_2$O ice band.

\noindent
{\bf GL 2136}-- The H$_2$O ice band optical depth toward this source
has been recently measured to be $\tau(3.1)=3.2\pm0.2$ (Kastner \&
Weintraub 1996).  The H$_2$O ice is partially crystalline (Fig. 1b).
The crystalline ice inflection at 3.20 $\mu$m precludes an accurate
test for the 3.25 $\mu$m feature with our current technique.

\noindent
{\bf SVS20}-- SVS20 is a double source in the Serpens cloud, with a
mostly N--S separation of $1.6^{\prime\prime}$ (Eiroa {\it et al.}
1987).  Our observations are of SVS20(S), which is 3 times brighter at
3.6 $\mu$m (Huard, Weintraub \& Kastner 1997).  Both objects exhibit
H$_2$O ice absorption bands (Eiroa \& Leinert 1987; Huard, Weintraub
\& Kastner 1997).

\noindent
{\bf RCrA/IRS1}--Graham (1998) estimated a 3.47 $\mu$m feature optical
depth of $\approx0.06$.  Our value is $0.041\pm0.005$, not remarkably
different.

\noindent
{\bf RCrA/IRS2}--Graham (1998) also estimated a 3.47 $\mu$m feature
optical depth of $\approx0.06$ for this source (TS 13.1).  Our value is
$0.041\pm0.011$, not remarkably different.

\noindent
{\bf GL 2591} -- Although this source is relatively bright, our
spectrum suffers from a poor cancellation of telluric features.
No attempt was made to fit a continuum to the 3.25 $\mu$m region.

\vskip 24pt
  
\section
{The 3.47 $\mu$m Feature}
 
\vskip 12pt

Using all of the available data, the best estimates of the central
wavelength and width of this feature are $\lambda_0=3.469\pm0.002$
$\mu$m (2883 cm$^{-1}$) and FWHM=$0.105\pm0.004$ $\mu$m (87
cm$^{-1}$).  Table 4 summarizes all of the detections of the feature
to date.  Figures 8a and 8b demonstrate that the 3.47 $\mu$m absorber
is better correlated with the 3.1 $\mu$m H$_2$O ice band depth, than
with the 9.7 $\mu$m silicate absorption band depth.  References for
the H$_2$O ice and silicate optical depths are given in Table 4.  This
is the same comparison done by Brooke, Sellgren, \& Smith (1996), but
the number of data points is now roughly doubled.  The best weighted
linear fit, exclusive of the upper limit, is
$$
\tau(3.47) = (0.033 \pm 0.002) \ \tau(3.1) - (0.004 \pm 0.004)   \eqno(1)
$$ 
which passes near the origin.  These coefficients are consistent with
the values derived by Brooke, Sellgren, \& Smith from the more limited
data.  The linear correlation coefficient is $r=0.88$.  The weighted
fit is especially sensitive to the optical depth of GL961E (open
circle at $\tau(3.1)=2.46$ in Fig. 8b), as its error as estimated by
Brooke, Sellgren, \& Smith is significantly lower than the points
around it.  In the event that there may be some as yet undetermined
systematic effects which affect the error determinations, we also
give the unweighted fit:
$$
\tau(3.47) = 0.042 \ \tau(3.1) - 0.010                            \eqno(2)
$$ 
In any case, the abundance of the 3.47 $\mu$m absorber is closely
related to that of H$_2$O ice over a wide range in ice extinction.

If in fact the 3.47 $\mu$m feature is due to the stretching vibration
of $sp^3$ C--H bonds, the linear relation of Fig. 8b corresponds to
roughly one C--H bond for every two H$_2$O molecules, and a possible
explanation for the correlation is that both H$_2$O ice and the C--H
bonds form in step on grains by hydrogen-addition reactions (Brooke,
Sellgren, \& Smith 1996).  Detections of longer-wavelength C--H
vibrational modes are needed to secure the identification; part of the
broad 6.8 $\mu$m absorption feature may be due to as yet unidentified
C--H deformation modes (Schutte {\it et al.} 1996b).

\vskip 24pt
\goodbreak
  
\section
{The 3.25 $\mu$m Feature}
 
\vskip 12pt

   The best estimates of the central wavelength and width of this
feature are $\lambda_0=3.250\pm0.004$ $\mu$m (3077 cm$^{-1}$) and
FWHM=$0.069\pm0.006$ $\mu$m (74 cm$^{-1}$).  Figures 9a and 9b show
the optical depth of the 3.25 $\mu$m feature compared to silicates and
H$_2$O ice.  With so few points, it is impossible to draw a firm
conclusion, but the data to date, including the upper limit, suggest a
better correlation with the silicates.  (The best linear fit with the
silicates has a nominal linear correlation coefficient $r=0.90$, but
with only 5 points, there is a $\approx5$ \% probability that the data
are completely uncorrelated.)

  If additional data show a correlation of the 3.25 $\mu$m feature
with the silicate feature, this would provide evidence to support the
Sellgren {\it et al.} (1995) identification of the feature as the C--H
stretch band of aromatic hydrocarbons at low temperature.  Aromatic
hydrocarbons are thought to be ubiquitous in the interstellar medium
and should correlate better with the refractory dust in molecular
clouds than with volatile ices.   

   The C--H stretch absorption band of aromatic hydrocarbons should
be accompanied by absorption counterparts of the well-known aromatic
emission features near 6.2, 7.7, 8.6, and 11.3 $\mu$m.  A possible
absorption feature at 6.25 $\mu$m toward the young stellar object
NGC7538/IRS9 has been suggested to be the C--C stretching mode of
aromatic hydrocarbons (Schutte {\it et al.} 1996b).  However, the
identification and optical depth are still uncertain due to the
presence of other absorbers.  There are as yet no published
high-sensitivity spectra of this object in the 3.25 $\mu$m region.

   A test of the possible correlation of the 3.25 $\mu$m feature
with the refractory dust would be the clear detection of the feature
along a line of sight with heavy dust extinction but little H$_2$O
ice, as can be found in the diffuse interstellar medium (ISM).  Such a
detection has not yet been made.  Previous searches for the aromatic
C--H stretch absorption in the diffuse ISM have been made near 3.29
$\mu$m, the wavelength of the aromatic emission feature.  Pendleton
{\it et al.} (1994) discussed a possible absorption feature at 3.28
$\mu$m, perhaps due to aromatic hydrocarbons, toward three Galactic
Center sources whose extinction is believed to be dominated by dust in
the diffuse ISM.  But the definition of the continuua was difficult,
so they placed upper limits of $\tau$(3.28) $\la$ 0.02 on this
feature.  Schutte et al. (1998) also placed an upper limit of
$\tau$(3.3) $\la$ 0.02 toward one Wolf-Rayet star in their sample of
diffuse ISM sources.  Sellgren {\it et al.} (1995) proposed that the
3.25 $\mu$m feature observed in absorption in molecular clouds may be
due to the same aromatic hydrocarbons responsible for the 3.29 $\mu$m
emission feature, but shifted to a shorter wavelength due to a
difference in temperature between cold absorbers and hot emitters.
High-sensitivity spectra of diffuse ISM sources in the 3.25 $\mu$m
region are needed to test this proposal.

   However an absorption feature at 6.2 $\mu$m identified as the
aromatic C--C stretch has recently been identified toward 5 Wolf-Rayet
stars and 2 Galactic Center sources in Infrared Space Obsevatory (ISO)
spectra (Schutte {\it et al.} 1998).  The extinction to all of these
objects is believed to come primarily from the diffuse ISM.  The
optical depth of the 6.2 $\mu$m feature is well-correlated with the
9.7 $\mu$m silicate absorption optical depth toward these sources.  In
Figure 10, we compare the relationships of the possible aromatic
absorptions at 6.2 and 3.25 $\mu$m and silicate absorption.  For this
comparison, we multiplied the $\tau$(6.2) values of Schutte {\it et
al.} by 0.25, the factor needed to bring the best-fit linear slopes
into agreement.  Figure 10 suggests that the apparent relationship
between $\tau$(3.25) and $\tau$(9.7) is consistent with the linear
correlation of $\tau$(6.2) vs. $\tau$(9.7), after correcting by a
scale factor.  This agreement is what one would expect if the
abundance of aromatic hydrocarbons relative to silicates, and the
relative strengths of C--H stretch and C--C stretch modes, were
similar in both molecular clouds and the diffuse ISM.

    Assuming that the 3.25 $\mu$m feature is due to aromatic
hydrocarbons, abundances can be estimated following Sellgren {\it et
al.}  (1995).  The abundances of aromatic C--H bonds relative to
hydrogen gas lie in the range N(C--H)/N(H)$\approx1--3 \times 10^{-5}$
(see Table 5).  Adopting a value for the ratio of aromatic carbon
atoms in hydrogen bonds to total aromatic carbon atoms of $f=0.4$
(Sellgren {\it et al.}  1995), roughly 8--23 \% of cosmic carbon would
be in aromatic hydrocarbons.

\vskip 12pt

\section
{Solid CH$_3$OH}

\vskip 12pt

The 3.53 $\mu$m feature due to solid CH$_3$OH was detected toward
three objects here: W33A, GL2136, and MonR2/IRS2.  The detection
toward MonR2/IRS2 is new.  A previous spectrum of NGC7538/IRS9 showing
solid CH$_3$OH (Brooke, Sellgren, \& Smith 1996) is also analyzed in
detail here.  For consistency with the other sources, the optical
depth profile of NGC7538/IRS9 in the 3.47 $\mu$m region was re-derived
using the same definition of the local continuum.  This required some
additional data at the long-wavelength end; these points were obtained
from Allamandola {\it et al.}  (1992) and were corrected for a slight
slope difference.  The derived gaussian parameters are given in Table
2.  They are not significantly different from those derived earlier.
We thus present analyses of four sources in which the 3.53 $\mu$m
feature of solid CH$_3$OH has been firmly identified to date.  The
feature has also recently been detected toward a fifth source,
RAFGL7009S (Dartois {\it et al.} 1998).

The peak wavelength of the CH$_3$OH $\nu_3$ band at 3.53 $\mu$m can be
used to constrain the composition of the methanol-containing ices
(Schutte {\it et al.}  1996a).  In mixtures with H$_2$O ice, the
feature shifts to shorter wavelengths (higher frequency) with
increasing H$_2$O fraction.  We used laboratory spectra of pure
CH$_3$OH and a 100:10:1:1 H$_2$O:CH$_3$OH:CO:NH$_3$ mix (``Weak
Interstellar Mix''), both at 10K, obtained in digital form from the
AAS CD-ROM version of Hudgins {\it et al.}  (1993).  The peak in pure
CH$_3$OH occurs at 3.535 $\mu$m (2829 cm$^{-1}$).  In the weak ISM
mix, it occurs at 3.530 $\mu$m (2833 cm$^{-1}$).

In order to fit the spectra, local continuua similar to those used for
the observations were applied to the CH$_3$OH laboratory data.  Each
of the spectra in Fig. 11 was simultaneously fitted with a gaussian for
the 3.47 $\mu$m feature and a CH$_3$OH ice template.  Optical depths
are quoted for the 3.53 $\mu$m feature alone.  An additional narrow
absorption feature near 3.47 $\mu$m in W33A discussed below was
deleted from the fit for this object.

The features in GL2136 and NGC7538/IRS9 occur at 3.536 $\mu$m (2828
cm$^{-1}$), in good agreement with pure CH$_3$OH ice.  In W33A, the
feature occurs at 3.534 $\mu$m (2830 cm$^{-1}$).  We have fit this
feature with a mix of the two samples described above, as indicated in
Table 6.

The feature in MonR2/IRS2 occurs at 3.528 $\mu$m (2834 cm$^{-1}$).
This is too short a wavelength to be consistent with pure CH$_3$OH.
The weak ISM mix provides a better, though still not exact, fit with
$\tau(3.53)=0.033\pm0.010$.  Presumably an even more water-rich mix
would improve the fit.  It would be desirable to have laboratory
spectra of such mixtures.

Previous studies indicated that the solid methanol toward W33A, GL
2136, and NGC7538/IRS9 is in a CH$_3$OH-rich phase with
CH$_3$OH/H$_2$O $\ga0.5$ (Skinner {\it et al.} 1992; Allamandola {\it
et al.} 1992; Schutte {\it et al.}  1996a; Dartois {\it et al.} 1998)
and our fits confirm this.  The 3.53 $\mu$m feature in MonR2/IRS2
suggests that solid methanol can also be found within H$_2$O-rich ice
having CH$_3$OH/H$_2$O$\la0.1$.

An additional indicator of CH$_3$OH in H$_2$O-rich environments would
be the presence of two weak features which appear at 3.38 $\mu$m and
3.41 $\mu$m (Schutte {\it et al.} 1996a).  These can be seen in the
laboratory data in Fig. 11.  The present signal-to-noise is not high
enough to constrain the presence of these features in any of the
sources.

Column densities and CH$_3$OH abundances are listed in Table 6.  For
W33A, the column density of H$_2$O ice is uncertain.  The range given
results from converting the observed optical depths of the 3.1 and 6.0
$\mu$m features to column densities.  But, as discussed above, the 3.1
$\mu$m band is most likely filled in by light which undergoes less ice
extinction.  On the other hand, the 6.0 $\mu$m band may contain other
absorbers (Schutte {\it et al.} 1996b).  The range given should
bracket the true line-of-sight column density.  It should be noted
that the CH$_3$OH column density may be an underestimate if the
filling-in noted above is also important at 3.53 $\mu$m.

In the other three sources, the CH$_3$OH/H$_2$O ice abundance derived
from the 3.53 $\mu$m band ranges from 3 to 5 \%.  The derived value of
the abundance for GL 2136 is in agreement with that derived by
Schutte {\it et al.}  (1996a) from the same band.  The derived values
for W33A and NGC7538/IRS9 are roughly a factor 3 lower than those of
Allamandola {\it et al.} (1992), due to the use of different band
strengths, different adopted continuua, and different assumed
contributions of the 3.47 $\mu$m feature.  Our abundance range for
W33A, 3--16 \% is in reasonable agreement with that of Dartois {\it et
al.} (1998), 5--22 \%.

   Upper limits to the CH$_3$OH/H$_2$O ice ratio toward other sources
have been estimated (Chiar, Adamson, \& Whittet 1996; Dartois {\it et
al.} 1998).  The most stringent is that for W3/IRS5 for which the upper
limit is of order 1 \%.  In contrast, Dartois {\it et al.} estimate a
very high abundance of 30 \% for RAFGL7009S.  Thus solid CH$_3$OH is
common along molecular cloud lines-of-sight with H$_2$O ice, with a
typical abundance of a few percent, but its abundance definitely
varies.

There are two notable aspects about the methanol abundances.  First,
the typical values (1--5 \% relative to H$_2$O) of the CH$_3$OH
abundance are very similar to the values in comets, which are derived
from the same band in gas-phase fluorescence (Hoban {\it et al.} 1991;
Bockel\'ee-Morvan, Brooke, \& Crovisier 1995; DiSanti {\it et al.}
1995).  This provides some supporting evidence for the direct
incorporation of interstellar ice grains into comets.

Second, the CH$_3$OH/H$_2$O ice ratio toward NGC7538/IRS9 is similar
to the ratios toward GL 2136 and MonR2/IRS2, but the relative
abundance of solid CO trapped in non-polar ices toward this source is
$\ga5$ times higher (Tielens {\it et al.} 1991; Chiar {\it et al.}
1998).  Thus the abundance of solid CH$_3$OH is not tightly coupled to
this CO component, which is believed to form in dense regions of
molecular clouds where most of the hydrogen is molecular (Tielens {\it
et al.} 1991).  This favors theories which postulate the formation of
solid CH$_3$OH by hydrogen-addition reactions in regions where atomic
hydrogen predominates, as has also been proposed for H$_2$O ice
(e.g. d'Hendecourt, Allamandola, \& Greenberg 1985).  The abundance of
solid CO trapped in polar ices (presumably H$_2$O) is also similar in
all three sources (2-3 \%, Tielens {\it et al.}  1991) and this CO
component may form under similar conditions.

\vskip 12pt

\section
{Additional Narrow Absorption Near 3.47 $\mu$m}

\vskip 12pt

There is an additional narrow absorption feature near 3.47 $\mu$m
(2882 cm$^{-1}$) in the spectrum of W33A (Fig. 11).  This region was
excluded from the fit with the 3.47 $\mu$m gaussian plus solid
methanol for W33A.  It is unlikely to be simply structure within the
broad 3.47 $\mu$m feature as it does not appear at the same level in
well-measured 3.47 $\mu$m features like those toward MonR2/IRS3
(Sellgren {\it et al.}  1995) and W3/IRS5 (Brooke, Sellgren, \& Smith
1996).  The feature is detected at the 3$\sigma$ level.  A similar
absorption feature was noted by Dartois {\it et al.} (1998).

Before examining the feature toward W33A, we discuss the spectrum of
GL 2136 in this region.  Schutte {\it et al.} (1996a) proposed that a
feature at 3.473 $\mu$m (2879 cm$^{-1}$) with FWHM$\approx25$
cm$^{-1}$ was present in GL 2136 with an optical depth
$\tau\approx0.01$.  They identified the feature as the $\nu_5$ band of
solid H$_2$CO in a mixture with H$_2$O and CH$_3$OH.  Our spectrum of
GL 2136 does not show the same structure which led Schutte {\it et
al.} to claim the presence of a feature, and the current spectrum has
higher signal-to-noise.  In our spectrum, this region resembles the
peak of a broad 3.47 $\mu$m absorption feature like that seen in other
sources.  However due to uncertainties in baselines we cannot rule out
some contribution from solid H$_2$CO at the bottom of the band.  Our
$3\sigma$ upper limit is $\tau\la0.01$, i.e. comparable to the claimed
detection, which corresponds to a solid H$_2$CO column density of
$\la3 \times 10^{17}$ cm$^{-2}$ and an abundance of $\la6$ \% relative
to H$_2$O using the same band width and strength adopted by Schutte
{\it et al.}  If this much H$_2$CO were present, part of the 3.53
$\mu$m band would be due to the $\nu_1$ H$_2$CO band and the solid
CH$_3$OH abundance would be roughly 9 \% lower.  But as we see no
evidence for solid H$_2$CO, we prefer attributing all of the 3.53
$\mu$m band to solid CH$_3$OH.

The narrow 3.47 $\mu$m feature in W33A falls at a wavelength
consistent with solid H$_2$CO in various mixtures with CH$_3$OH and
H$_2$O (Schutte {\it et al.} 1996a), but the feature appears
asymmetric, unlike H$_2$CO.  If the excess absorption is interpreted
as entirely due to the $\nu_5$ band of H$_2$CO, the peak optical depth
of $\tau\approx0.06$ would imply a column density of $1.8 \times
10^{18}$ cm$^{-2}$.  As discussed above, the H$_2$O ice column density
to W33A is uncertain, but this upper limit to the optical depth
corresponds to a solid H$_2$CO abundance of $\la$4-20 \% relative to
H$_2$O.  If this feature were due to H$_2$CO, the $\nu_1$ H$_2$CO band
would contribute to the 3.53 $\mu$m band and methanol abundances would
be $\approx12$ \% lower.

Another possible contributor to the narrow 3.47 $\mu$m feature is
solid ethane, C$_2$H$_6$, which has a narrow (FWHM$\approx6$
cm$^{-1}$) band at 3.472 $\mu$m (2880 cm$^{-1}$) (Boudin, Schutte, \&
Greenberg 1998).  A stronger C$_2$H$_6$ band occurs at 3.365 $\mu$m
(2972 cm$^{-1}$) and the optical depth of this feature should be 2-3
times higher.  The present data for W33A would be consistent with a
3$\sigma$ optical depth upper limit of approximately 0.06 for this
feature.  Thus up to about half of the apparent narrow 3.47 $\mu$m
feature could be due to solid C$_2$H$_6$.  This would correspond to a
solid C$_2$H$_6$ column density of $3.0 \times 10^{16}$ cm$^{-2}$ or
0.1-0.3 \% relative to H$_2$O ice.  This is near the abundance of
C$_2$H$_6$ in comets C/1996 B2 Hyakutake and C/1995 O1 Hale-Bopp of
$\sim$0.4 \% relative to H$_2$O (Mumma {\it et al.} 1996; Weaver {\it
et al.}  1998).  Future, higher signal-to-noise spectra of W33A in
this region should be able to distinguish possible contributions of
solid H$_2$CO and C$_2$H$_6$.

%
%
%
%

\vskip 12pt

\section
{Conclusions}

\vskip 12pt

\noindent
1) The optical depth of the broad absorption feature at 3.47 $\mu$m
seen on the long-wavelength wing of molecular cloud H$_2$O ice bands
is correlated with the optical depth of H$_2$O ice, in agreement with
an earlier conclusion based on a smaller data set.  If the feature is
due to C--H bonds, the abundance of the C--H bonds must be closely
related to that of H$_2$O ice.

\noindent
2) The broad absorption feature at 3.25 $\mu$m was detected toward two
more young stellar objects.  There are insufficient data to draw any
firm conclusions, but the optical depths measured to date correlate
better with the refractory silicates than with H$_2$O ice, providing
evidence that supports an identification with aromatic hydrocarbons at
low temperature.

\noindent
3) The 3.53 $\mu$m solid CH$_3$OH features toward W33A, GL2136, and
NGC7538/IRS9 can be fit by CH$_3$OH-rich ices, while the wavelength of
the feature toward MonR2/IRS2 suggests an H$_2$O-rich ice environment.
Solid methanol abundances toward GL 2136, NGC7538/IRS9, and MonR2/IRS2
are 3-5 \% relative to H$_2$O ice, similar to cometary abundances.

\noindent
4) There is an additional narrow absorption feature near 3.47 $\mu$m
toward W33A.  Possible contributors are solid H$_2$CO and solid
C$_2$H$_6$.

\noindent
5) The H$_2$O ice toward W51/IRS2 is located primarily in front of the
eastern component, is partially crystalline, and has an optical depth
$\tau(3.1)=2.4\pm0.2$.  Spectra of RNO 91 from 2 to 4 $\mu$m reveal
stellar (or circumstellar) CO gas absorption and an H$_2$O ice band
optical depth $\tau(3.1)=2.1\pm0.2$.

\vskip 12pt
 
\acknowledgments
{ACKNOWLEDGMENTS}
\vskip 12pt

The spectra from September 1995 were obtained through the UKIRT
Service Programme.  R. Joyce (NOAO) obtained and reduced the spectrum
of MonR2/IRS2 for us.  We thank D. Crisp (JPL) for making computer
resources available.

\vskip 12pt

\clearpage
\begin{table}
\begin{center}
\begin{tabular}{lrccrcrr}
\multicolumn{8}{c}{{\bf Table 1: Log of Observations}}\\
\hline\hline
\multicolumn{1}{c}{Object} & \multicolumn{1}{c}{UT} & Tel & Range & \multicolumn{1}{c}{t$^a$} & Standard & \multicolumn{1}{c}{Spectral} & \multicolumn{1}{c}{${T_{\rm BB}}^b$} \\
      &     &     &($\mu$m)&\multicolumn{1}{c}{(sec)} & Star      &\multicolumn{1}{c}{Type}    &\multicolumn{1}{c}{(K)}     \\
\hline

MonR2/IRS2 & 12 Jan 1996 & 2.1m & 3.3-3.6 & 180 & $\epsilon$ Ori & B0I & 26000 \\
$\rho$ Oph/Elias 29   & 12 Jun 1996 & UKIRT & 3.1-3.4 & 288 & 49 Lib      & F8V & 6200 \\  
           & 13 Jun 1996 &       & 3.3-3.6 & 288 & BS 5923     & F6V & 6400 \\
           & 16 Jul 1996 &       & 2.9-3.2 & 288 & $\delta$ Sco& B0IV & 27000 \\
$\rho$ Oph/WL 6       & 13 Jun 1996 & UKIRT & 3.1-3.4 & 432 & BS 6496     & F7V & 6300 \\
           & 13 Jun 1996 &       & 3.3-3.6 & 432 & BS 6310     & F3V & 6700 \\
           & 16 Jul 1996 &       & 2.9-3.2 & 432 & $\delta$ Sco& B0IV & 27000 \\ 
$\rho$ Oph/Elias 21   & 12 Jun 1996 & UKIRT & 3.1-3.4 & 288 & BS 6310     & F3V & 6700 \\
           & 12 Jun 1996 &       & 3.3-3.6 & 288 & BS 6310     & F3V & 6700 \\
           & 16 Jul 1996 &       & 2.9-3.2 & 432 & $\delta$ Sco& B0IV & 27000 \\
$\rho$ Oph/Elias 33   & 12 Jun 1996 & UKIRT & 3.1-3.4 & 288 & BS 6310     & F3V & 6700 \\
           & 12 Jun 1996 &       & 3.3-3.6 & 288 & BS 6310     & F3V & 6700 \\
           & 16 Jul 1996 &       & 2.9-3.2 & 432 & BS 7152     & F0V & 7200 \\
RNO 91     & 06 Jun 1996 & 2.1m & 2.0-2.4 & 480 & $\nu$ Ser & A2V & 9000 \\
           & 06 Jun 1996 &      & 3.0-3.9 & 240 & $\nu$ Ser & A2V &9000 \\
           & 13 Jun 1996 & UKIRT& 3.1-3.4 & 432 & BS 6496 & F7V & 6300 \\
           & 13 Jun 1996 &      & 3.3-3.6 & 432 & BS 6012 & F3V & 6700 \\
           & 16 Jul 1996 &      & 2.9-3.2 & 432 & $\delta$ Sco& B0IV &
27000 \\
W33A       & 06 Sep 1995 & UKIRT & 2.9-3.6 & 1440 & BS 6496   & F7V &
6300 \\
GL 2136     & 06 Sep 1995 & UKIRT & 2.9-3.6 & 576 & BS 7126 & F5V & 6440 \\   
Ser/SVS20  & 04 Jun 1996 & 2.1m  & 3.3-3.6 &180 & $\delta$ Aql & F2IV & 6890 \\
           & 16 Jul 1996 & UKIRT & 2.9-3.2 & 240 & BS 6797     & F5V & 6440 \\
           & 16 Jul 1996 &       & 3.1-3.4 & 432 & BS 6797     & F5V & 6440 \\
RCrA/IRS1  & 12 Jun 1996 & UKIRT & 3.3-3.6 & 288 & BS 7152     & F0V & 7200 \\
           & 13 Jun 1996 &       & 3.1-3.4 & 432 & BS 7152     & F0V & 7200 \\
           & 16 Jul 1996 &       & 2.9-3.2 & 288 &  BS 7152     & F0V & 7200 \\
RCrA/IRS2  & 13 Jun 1996 & UKIRT & 3.1-3.4 & 432 & BS 7152     & F0V & 7200 \\
           & 13 Jun 1996 &       & 3.3-3.6 & 432 &  BS 7152     & F0V & 7200 \\
           & 16 Jul 1996 &       & 2.9-3.2 & 240 & $\delta$ Sco& B0IV & 27000 \\
RCrA/IRS5  & 13 Jun 1996 & UKIRT & 3.3-3.6 & 432 & BS 7152     & F0V & 7200 \\
W51/IRS2   & 06 Jun 1996 & 2.1m  & 2.0-2.4 &120 & 15 Vul      & A7m & 7850 \\
           & 06 Jun 1996 &       & 3.0-3.9 &320 & 111 Her     & A3V & 8720 \\
           & 12 Jun 1996 & UKIRT & 3.1-3.4 & 288 & BS 7172     & F8V & 6200 \\
           & 13 Jun 1996 &       & 2.9-3.2 & 432 & BS 7550     & F5V & 6440 \\
GL 2591    & 06 Sep 1995 & UKIRT & 2.9-3.6 & 576 & $\tau$ Cyg& F3IV & 6700 \\
\hline\hline

\end{tabular}
\end{center}
\end{table}
 
Notes to Table 1: 
 
{$^a$} Integration time

{$^b$} Assumed blackbody temperature of standard star. 

\clearpage
\begin{table}
\begin{center}
\begin{tabular}{lccc}
\multicolumn{4}{c}{{\bf Table 2: 3.47 $\mu$m Absorption Feature Parameters$^{ab}$}}\\
\hline\hline
\multicolumn{1}{c}{Object} &${\lambda_0}$&$\Delta\lambda$&$\tau$\\
&($\mu$m)&($\mu$m)&\\
\hline
\\
MonR2/IRS2  & 3.460  & 0.103  & 0.083 \\
            &(0.006) &(0.007) &(0.013) \\
$\rho$ Oph/Elias 29    & 3.490  & 0.113  & 0.089 \\
            &(0.006) &(0.009) &(0.007) \\
$\rho$ Oph/WL 6        & 3.472  & 0.111  & 0.062 \\
            &(0.003) &(0.009) &(0.012) \\
$\rho$ Oph/Elias 21    & 3.464  & 0.104  & 0.042 \\
            &(0.003) &(0.012) &(0.012) \\
$\rho$ Oph/Elias 33    & 3.457  & 0.120  & 0.024 \\
            &(0.009) &(0.007) &(0.006) \\
RNO 91      & 3.469  & 0.118  & 0.085 \\
            &(0.003) &(0.012) &(0.015)\\
W33A        & 3.477  & 0.099  & 0.290$^c$ \\
            &(0.010) &(0.012) &(0.040)\\
GL 2136     & 3.471  & 0.118  & 0.137 \\
            &(0.010) &(0.005) &(0.020) \\
Ser/SVS20   & 3.463  & 0.111  & 0.030 \\
            &(0.011) &(0.009) &(0.009) \\
RCrA/IRS1   & 3.469  & 0.122  & 0.041 \\
            &(0.012) &(0.016) &(0.005) \\
RCrA/IRS2   & 3.472  & 0.092  & 0.041 \\
            &(0.005) &(0.005) &(0.011) \\
RCrA/IRS5   & 3.470  & 0.111  & 0.090 \\
            &(0.006) &(0.005) &(0.019) \\
W51/IRS2(E) & 3.471  & 0.094  & 0.143 \\
            &(0.003) &(0.005) &(0.012) \\
GL 2591     & 3.464  & 0.131  & 0.045 \\
            &(0.005) &(0.007) &(0.005) \\
NGC7538/IRS9 & 3.485 & 0.118 & 0.130 \\
             &(0.003)&(0.024)&(0.017) \\
\hline\hline

\end{tabular}
\end{center}
\end{table}

Notes to Table 2: 
 
{$^a$} Central wavelengths, full widths at half maximum, and
peak optical depths of absorption features from gaussian fits.

{$^b$} Entries in parentheses are 1$\sigma$ uncertainties obtained
from standard deviations of results using several different baselines
(see text).
 
{$^c$} Interpreted as a lower limit (see text).

\clearpage
\begin{table}
\begin{center}
\begin{tabular}{lccc}
\multicolumn{4}{c}{{\bf Table 3: 3.25 $\mu$m Absorption Feature Parameters$^{ab}$}}\\
\hline\hline
\multicolumn{1}{c}{Object} & ${\lambda_0}$&$\Delta\lambda$&$\tau$\\
&($\mu$m)&($\mu$m)&\\
\hline
\\
Ser/SVS20   & 3.245  & 0.061  & 0.021 \\
            &(0.003) &(0.007) &(0.003) \\
RCrA/IRS1   & 3.239  & 0.059  & 0.032 \\
            &(0.003) &(0.012) &(0.010) \\
$\rho$ Oph/Elias 29 & &       & $<$0.03$^c$ \\
\hline\hline

\end{tabular}
\end{center}
\end{table}

Notes to Table 3: 
 
{$^a$} Central wavelengths, full widths at half maximum, and
peak optical depths of absorption features from gaussian fits.

{$^b$} Entries in parentheses are 1$\sigma$ uncertainties obtained
from standard deviations of results using several different baselines
(see text).

{$^c$} 3$\sigma$ upper limit.
 
\clearpage
\begin{table}
\begin{center}
\begin{tabular}{lcccc}
\multicolumn{5}{c}{{\bf Table 4: Summary of Absorption Feature Optical Depths}}\\
\hline\hline
\multicolumn{1}{c}{Object} & $\tau$(9.7) & $\tau$(3.1) & $\tau$(3.47) & $\tau$(3.25) \\[12pt]
&&&& \\
\hline
\\
W3/IRS5             & 7.64$^a$ & 3.48$^b$ & 0.13$^c$   & --        \\
Tau/Elias 18        & 0.43$^d$ & 0.80$^d$ & 0.022$^c$  & --        \\
HL Tau              & 0.56$^e$ & 0.77$^f$ & 0.035$^g$  & --        \\
Tau/Elias 16        & 0.66$^d$ & 1.26$^h$ & 0.032$^g$  & --        \\
BN                  & 3.28$^a$ & 1.78$^b$ & 0.034$^c$  & --        \\
Mon R2/IRS2         & --       & 2.54$^b$ & 0.083$^i$  & --        \\
Mon R2/IRS3         & 4.30$^a$ & 1.14$^b$ & 0.036$^j$  & 0.049$^j$ \\
S255/IRS1           & 5.11$^a$ & 1.48$^b$ & 0.024$^c$  & --         \\
GL961E              & 2.11$^a$ & 2.46$^b$ & 0.068$^c$  & --         \\
$\rho$ Oph/Elias 29 & 1.51$^k$ & 1.80$^l$ & 0.089$^i$  & $<$0.03$^i$    \\
$\rho$ Oph/WL 6     & 1.22$^k$ & 2.1$^l$  & 0.062$^i$  & --          \\
$\rho$ Oph/Elias 21 & 0.61$^k$ & 0.77$^l$ & 0.042$^i$  & --          \\
$\rho$ Oph/Elias 33 & 0.36$^k$ & 1.00$^l$ & 0.024$^i$  & --           \\  
RNO 91              & --       & 2.1$^i$  & 0.085$^i$  & --            \\
W33A                & 7.84$^a$ & $>$5.4$^a$ & $>$0.29$^i$ & --        \\
GL 2136             & 5.07$^a$ & 3.2$^m$  & 0.14$^i$   & --           \\
Ser/SVS20           & 2.0$^n$  & 1.0$^o$  & 0.030$^i$  & 0.021$^i$    \\
RCrA/IRS1           & 1.21$^p$ & 1.44$^p$ & 0.041$^i$  & 0.032$^i$     \\
RCrA/IRS2           & --       & 1.1$^q$  & 0.041$^i$  & --           \\
RCrA/IRS5           & --       & 2.3$^q$  & 0.090$^i$  & --           \\
W51/IRS2(E)         & 3.7$^r$  & 2.4$^i$  & 0.14$^i$  & --           \\
GL 2591             & 4.14$^a$ & 0.92$^b$ & 0.045$^i$  & --           \\
S140/IRS1           & 3.97$^a$ & 1.28$^b$ & 0.027$^c$  & 0.036$^c$    \\
NGC 7538/IRS1       & 6.38$^a$ & 1.29$^b$ & 0.052$^c$  & 0.078$^c$    \\
NGC 7538/IRS 9      & 4.46$^a$ & 3.28$^b$ & 0.13$^i$   & --           \\
\hline

\end{tabular}
\end{center}
\end{table}
 
Notes to Table 4: The optical depths given in the table are for
silicates at 9.7 $\mu$m; H$_2$O ice at 3.1 $\mu$m; and broad features
at 3.47 and 3.25 $\mu$m.  The silicate optical depth for W51/IRS2(E)
was corrected for intrinsic silicate emission following Willner {\it
et al.} 1982.
 
{$^a$} Willner {\it et al.} 1982.

{$^b$} Smith, Sellgren, \& Tokunaga 1989.

{$^c$} Brooke, Sellgren \& Smith 1996.

{$^d$} Whittet {\it et al.} 1988. 

{$^e$} Hanner, Brooke, \& Tokunaga 1998.

{$^f$} Sato {\it et al.} 1990.

{$^g$} Chiar, Adamson, \& Whittet 1996.

{$^h$} Smith, Sellgren, \& Brooke 1993.

{$^i$} This work.

{$^j$} Sellgren, Smith \& Brooke 1994.

{$^k$} Hanner, Brooke, \& Tokunaga 1995.

{$^l$} Tanaka {\it et al.} 1990.

{$^m$} Kastner \& Weintraub 1996.

{$^n$} J. Guertler, priv. comm.

{$^o$} Eiroa \& Hodapp 1989.

{$^p$} Whittet {\it et al.} 1996.

{$^q$} Tanaka {\it et al.} 1994.

{$^r$} Genzel {\it et al.} 1982.

\clearpage
\begin{table}
\begin{center}
\begin{tabular}{ccccc}
\multicolumn{5}{c}{{\bf Table 5: Summary of 3.25 $\mu$m Features}}\\[10pt]
\hline\hline

&$\tau$(3.25)$^a$& N(C--H)$^b$ & N(H)$^c$ & [C--H/H]$^d$ \\
&& (10$^{18}$ cm$^{-2}$) & (10$^{23}$ cm$^{-2}$) & \\
\hline
\\
Mon R2/IRS3 & 0.049   & 2.1   & 1.5 & 1.4 \\
            & (0.007) & (0.3) &     &(0.2) \\
\\
Elias 29    & $<0.03$ & $<1.3$& 0.53 & $<2.4$ \\
            &         &       &      &        \\
\\
SVS 20      & 0.021   & 0.91   & 0.70 & 1.3 \\
            & (0.003) & (0.13) &     &(0.2) \\
\\
RCrA/IRS1   & 0.032   & 1.4   & 0.42 & 3.3 \\
            & (0.010) & (0.4) &     &(0.9) \\
\\
S140/IRS1   & 0.036   & 1.6   & 1.4 & 1.1 \\
            & (0.007) & (0.3) &     &(0.2) \\
\\
NGC7538/IRS1 & 0.078   & 3.4   & 2.2 & 1.5 \\
             & (0.013) & (0.6) &     &(0.3) \\
\\
\hline\hline

\end{tabular}
\end{center}
\end{table}
 
Notes to Table 5:

{$^a$} Optical depth of 3.25 $\mu$m absorption feature.  1$\sigma$
error in parentheses.  References for $\tau(3.25)$ are in Table 4.

{$^b$} Column density of aromatic C--H bonds calculated from
N(C--H)=$\tau$(3.25)$\Delta\nu$/A with A=$1.7 \times 10^{-18}$ cm/C--H
bond (Sellgren {\it et al.} 1995) and $\Delta\nu=74$ cm$^{-1}$.
1$\sigma$ error in parentheses.

{$^c$} Column density of hydrogen gas estimated from N(H)=$3.5 \times
10^{22} \times \tau(9.7)$ cm$^{-2}$ (Tielens {\it et al.} 1991).
References for $\tau(9.7)$ are in Table 4.

{$^d$} Abundance of aromatic C--H bonds relative to hydrogen gas.  1$\sigma$
error in parentheses.

\clearpage
\begin{table}
\begin{center}
\begin{tabular}{cccccc}
\multicolumn{6}{c}{{\bf Table 6: Solid CH$_3$OH Detections at 3.53 $\mu$m}}\\[10pt]
\hline\hline

&$\tau$(3.53)$^a$& Fit Type$^b$ & N(CH$_3$OH)$^c$ & N(H$_2$O)$^d$ & [CH$_3$OH/H$_2$O]$^e$ \\
&&& (10$^{17}$ cm$^{-2}$) & (10$^{19}$ cm$^{-2}$) & \\
\hline
\\
Mon R2/IRS2 & 0.033 & Weak ISM Mix & 1.3 & 0.42$^f$ & 0.031 \\
            & (0.010) &            & (0.4) &        &(0.009) \\
\\
W33A        & 0.27  & Pure CH$_3$OH &11.9  & 0.9-4.2$^g$  & 0.03-0.16 \\
            &(0.05) &      +        &(2.2) &              &           \\
            & 0.07  & Weak ISM Mix  & 2.7  &      &        \\
            &(0.02) &               &(0.8) &      &        \\
\\
GL2136      & 0.060 & Pure CH$_3$OH & 2.6  & 0.50$^h$     & 0.052 \\
            &(0.010)&               &(0.4) &              &(0.009) \\
\\
NGC7538/IRS9 & 0.070    & Pure CH$_3$OH & 3.0 & 0.7$^i$   & 0.043 \\
             &(0.01)    &               &(0.4)&           &(0.006) \\
\\
\hline\hline

\end{tabular}
\end{center}
\end{table}
 
Notes to Table 6:

{$^a$} Optical depth of 3.53 $\mu$m absorption feature.  1$\sigma$
error in parentheses.

{$^b$} For the Weak ISM Mix, a FWHM $\Delta\nu$=29 cm$^{-1}$ and absorbance
A=$7.5 \times 10^{-18}$ cm/molecule (Hudgins {\it et al.} 1993) were
assumed.  For pure CH$_3$OH, a FWHM $\Delta\nu$=29 cm$^{-1}$ and absorbance
A=$6.6 \times 10^{-18}$ cm/molecule (Schutte {\it et al.} 1996a) were
assumed.

{$^c$} Column density of solid CH$_3$OH calculated from
N(CH$_3$OH)=$\tau$(3.53)$\Delta\nu$/A (Allamandola {\it et al.}
1992).  1$\sigma$ error in parentheses.

{$^d$} Column density of H$_2$O ice.

{$^e$} Abundance of solid CH$_3$OH relative to H$_2$O ice.  1$\sigma$
error in parentheses.

{$^f$} Estimated from the 3.1 $\mu$m band measured by Smith, Sellgren,
\& Tokunaga (1989).

{$^g$} Estimated from the 3.1 and 6.0 $\mu$m bands by Allamandola {\it
et al.} (1992).

{$^h$} Estimated from the 3.1 $\mu$m band by Schutte {\it et al.} (1996a).

{$^i$} Estimated from the 3.1 $\mu$m band by Allamandola {\it et al.} (1992).

\clearpage
\centerline
{\bf Figure Captions}
 
\vskip 12pt
 
{\bf Figure 1($ab$)}--- Spectra from 2.9 to 3.6 $\mu$m with resolution
$\lambda$/$\Delta \lambda \approx 1200$ (solid lines).  Dashed lines
are polynomial fits in local continuum regions (see text).  ($a$)
Filled circles are data from Allamandola {\it et al.} (1992) scaled to
the same flux level.  ($b$) Filled circles are data from Schutte {\it
et al.}  (1996a) scaled to the same flux level.  Apparent hydrogen
line emission features at 3.04 $\mu$m (HI Pf$\epsilon$) and 3.30
$\mu$m (HI Pf$\delta$) in GL 2136 may be due all or in part to
corresponding absorption features in the standard stars, which were
not corrected for.

{\bf Figure 2($a$-$h$)}--- Spectra from 2.9 to 3.6 $\mu$m with
effective resolution $\lambda$/$\Delta \lambda \approx 1300$ (1200 for
GL 2591) are shown as solid lines.  Dashed lines are polynomial fits
in local continuum regions (see text).  Apparent hydrogen line
emission features at 3.04 $\mu$m (HI Pf$\epsilon$) and 3.30 $\mu$m (HI
Pf$\delta$) in Elias 29, SVS20, RCrA/IRS1, and RCrA/IRS2 may be due
all or in part to corresponding absorption features in the standard
stars, which were not corrected for.

{\bf Figure 3($ab$)}--- Same as Figure 2 for objects observed only
between 3.3 and 3.6 $\mu$m. Filled circles for MonR2/IRS2 are data
from Smith, Sellgren, \& Tokunaga (1989) scaled to the same flux
level.

{\bf Figure 4($ab$)}--- 2-4 $\mu$m spectra. ($a$) For W51/IRS2(E),
spectral resolution was $\approx 700$ in the 2.2 $\mu$m region,
$\approx 1300$ in the 2.9-3.4 $\mu$m region and $\approx 530$ in the
3.4-3.9 $\mu$m region.  Long dashed line is estimated continuum.  A
scaled blackbody (dashed-dotted line) indicates the approximate color
temperature.  Short dashed line is the local continuum for the 3.47
$\mu$m feature.  Wavelengths of the indicated lines are (in microns):
2.059 (HeI 2p$^1$P$^0$-2s$^1$S), 2.113 (HeI 4s$^3$S-3p$^3$P$^0$),
2.122 (H$_2$ v=1-0 S(1)), 2.166 (HI Br$\gamma$), 3.039 (HI
Pf$\epsilon$), 3.297 (HI Pf$\delta$), 3.741(HI Pf$\gamma$).  After
correction for hydrogen absorption in the standard stars, the line
flux ratios (Wm$^{-2}$) relative to Br$\gamma$ of the indicated lines
are (in increasing wavelength order): 0.63, 0.04, 0.03, 1.0, 0.38,
0.91, 5.21.  ($b$) For RNO 91 the effective spectral resolution was
$\approx 400$ in the 2.2 $\mu$m region, $\approx 1300$ in the 2.9-3.6
$\mu$m region, and $\approx 200$ in the 3.6-3.9 $\mu$m region.  Long
dashed line is estimated continuum.  Short dashed line is the local
continuum for the 3.47 $\mu$m feature.  The wavelengths of the CO band
heads are (in microns): 2.293 (2-0), 2.323 (3-1), 2.352 (4-2), 2.383
(5-3).

{\bf Figure 5($ab$)}---Derived H$_2$O ice band profiles from the
continuua indicated in Figure 4.  Wavelengths of the hydrogen emission
lines are given in Fig. 4.

{\bf Figure 6($a$-$h$)}--- Derived 3.47 $\mu$m feature optical depth
profiles from continuua indicated in Figs. 2-3.  Dashed lines are
gaussian fits.  A representative $\pm1\sigma$ error bar is offset lower
right.

{\bf Figure 7($a$-$c$)}--- Derived 3.25 $\mu$m and 3.47 $\mu$m feature
optical depth profiles from continuua indicated in Fig. 2.  Dashed
lines are gaussian fits.  Some $\pm1\sigma$ error bars are indicated.
 
{\bf Figure 8($ab$)}---Optical depths for the 3.47 $\mu$m feature
vs. ($a$) silicates and ($b$) H$_2$O ice.  Solid line is the best
weighted linear fit.  Dashed line is best unweighted linear fit.  Open
circles are from Sellgren, Smith, \& Brooke (1994) and Brooke,
Sellgren, \& Smith (1996); X's from Chiar, Adamson, \& Whittet (1996);
filled circles from this work.

{\bf Figure 9($ab$)}---Optical depths for the 3.25 $\mu$m feature
vs. ($a$) silicates and ($b$) H$_2$O ice.  Open circles are from
Sellgren, Smith, \& Brooke (1994) and Brooke, Sellgren, \& Smith
(1996); filled circles from this work.

{\bf Figure 10}---Optical depths of the 3.25 $\mu$m absorption feature
measured toward molecular cloud sources, and scaled values for the 6.2
$\mu$m absorption feature measured toward sources whose extinction is
believed to come primarily from the diffuse ISM (Schutte {\it et al.}
1998), compared to silicate absorption optical depths.  Open circles
are from Sellgren, Smith, \& Brooke (1994) and Brooke, Sellgren, \&
Smith (1996); filled circles from this work.  The Schutte {\it et al.}
sources are 5 Wolf-Rayet stars (WR118, WR112, WR104, WR98a, WR48a) and
an upper limit for Cyg OB2 No. 12 (small filled squares); and 2
Galactic Center sources, GCS3 and GCS4 (small filled diamonds).  The
scaling factor of 0.25 brings the best-fit linear slopes into
agreement.

{\bf Figure 11}--- Derived optical depths from continuum fits for
sources with the 3.53 $\mu$m feature.  Short dashed line is the best
fit consisting of the sum of a 3.47 $\mu$m feature gaussian (long
dashes) and solid CH$_3$OH (dotted-dashed) as described in text.  Some
representative $\pm1\sigma$ error bars are indicated.

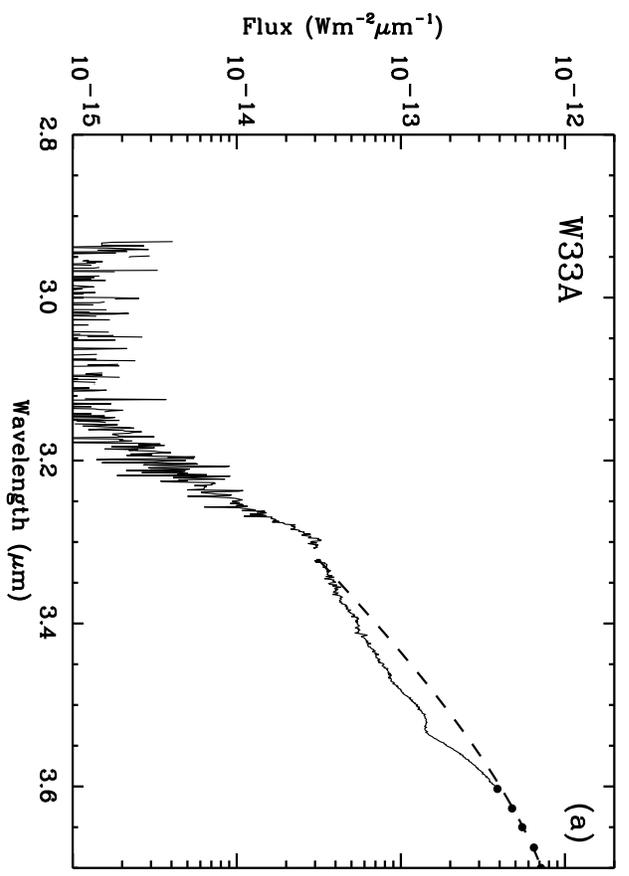
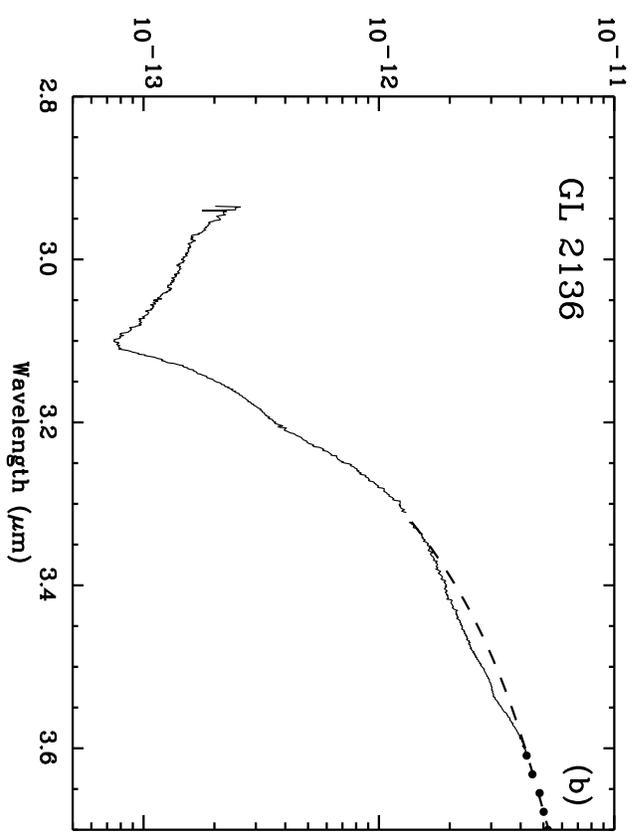
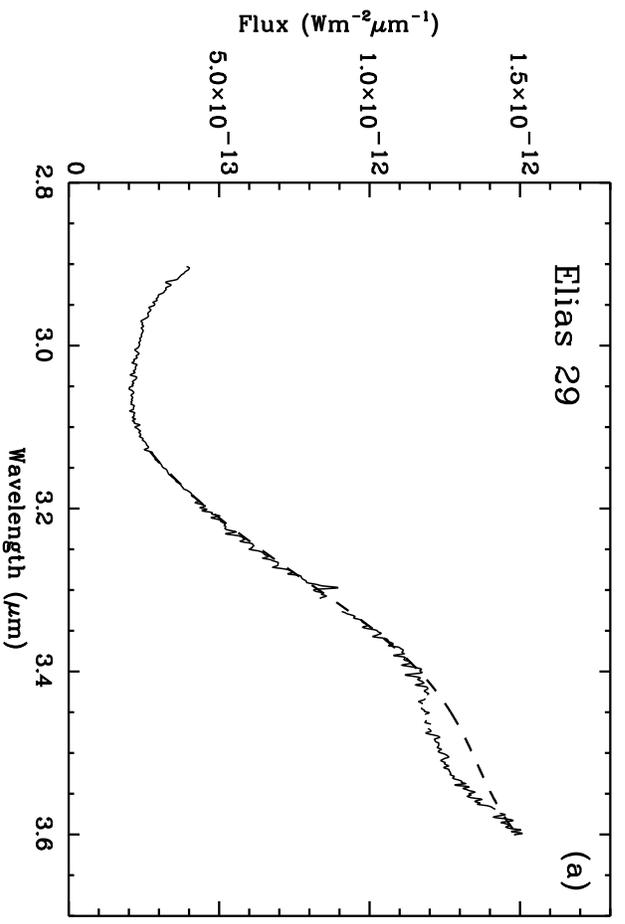
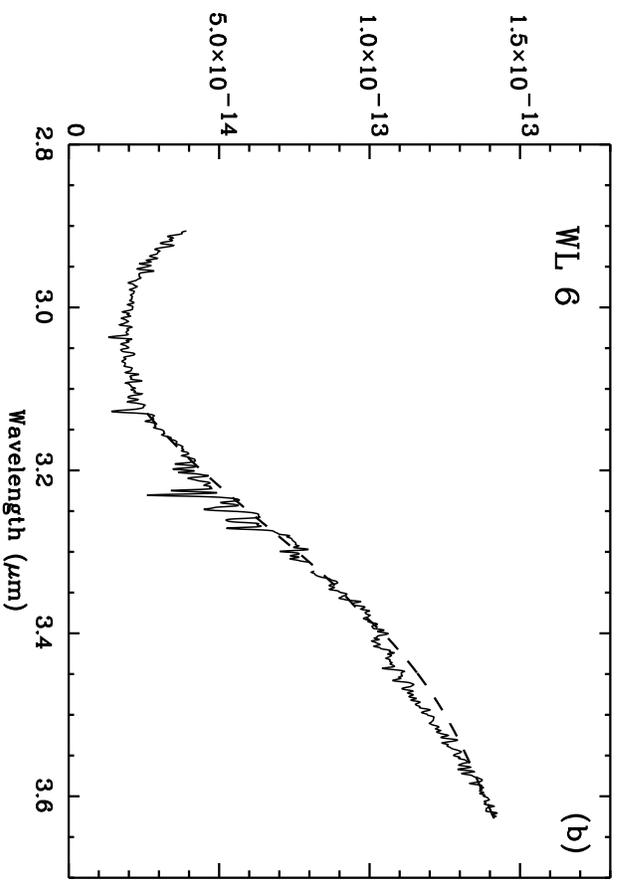
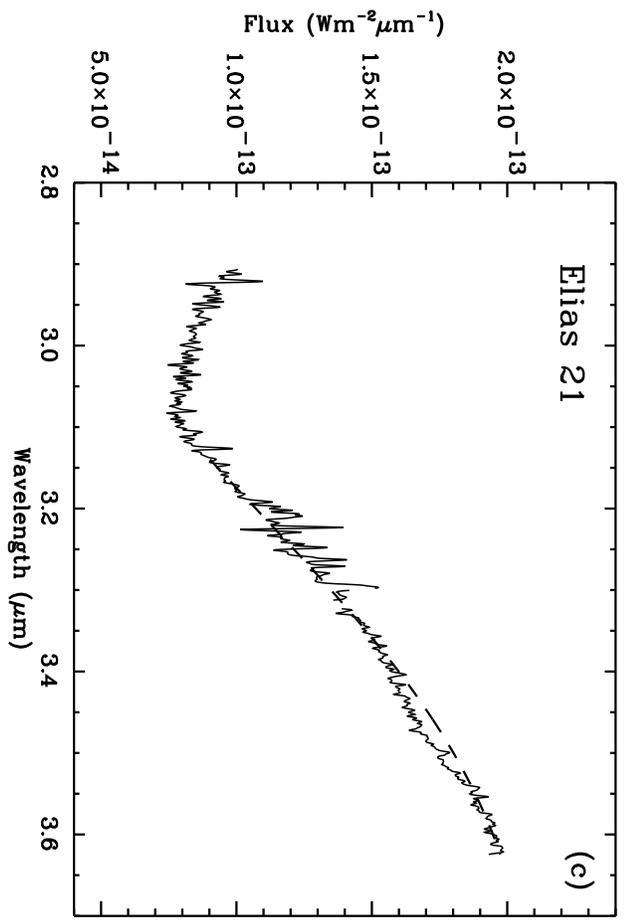
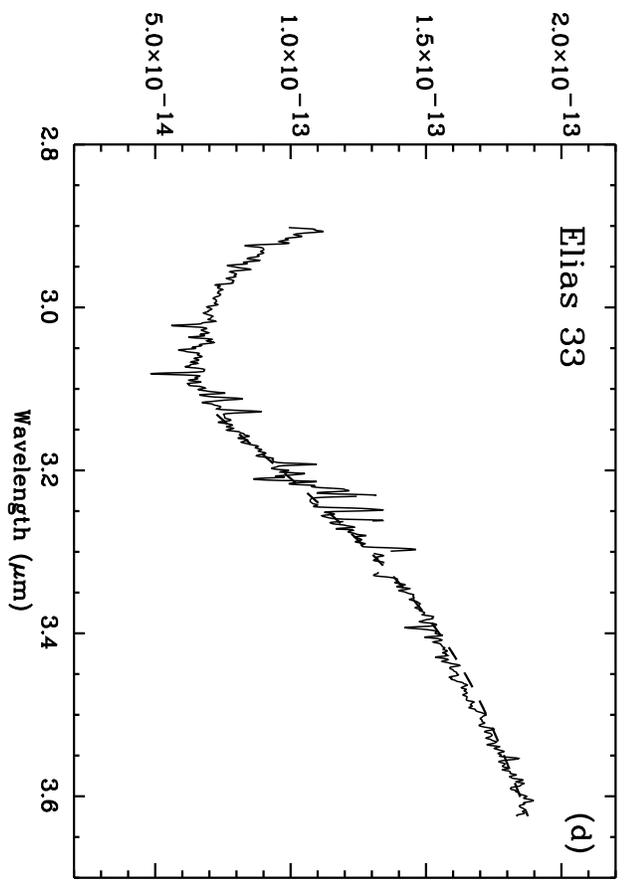
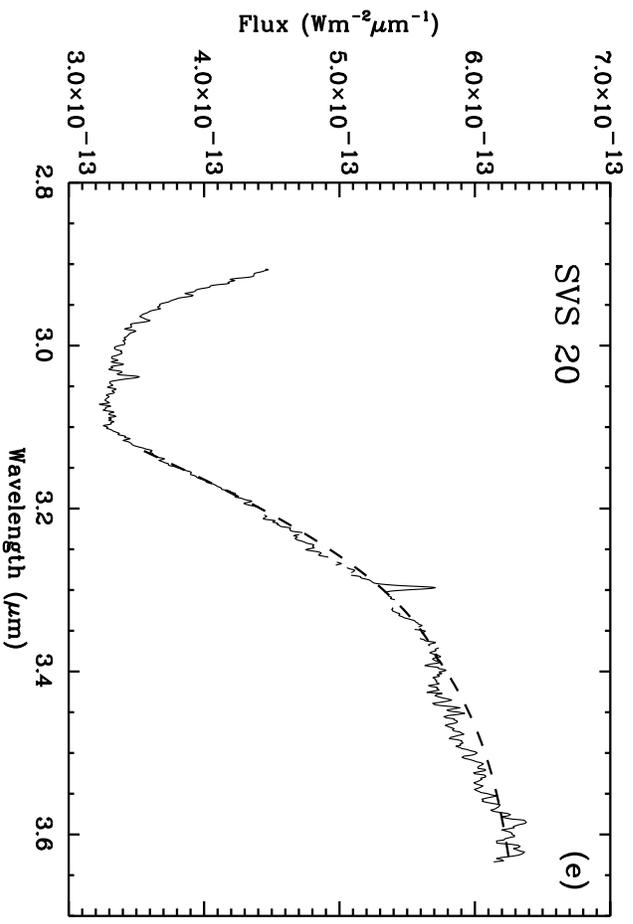
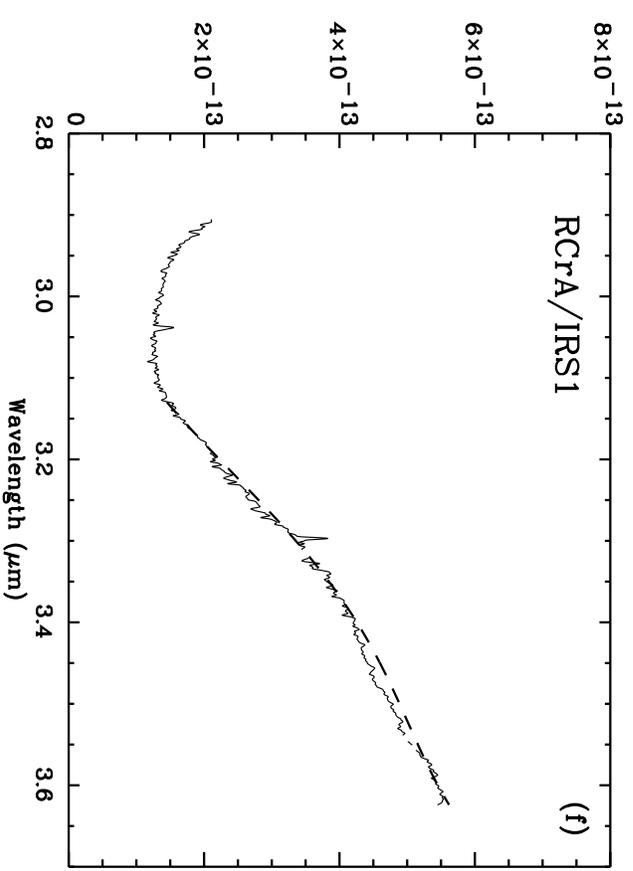
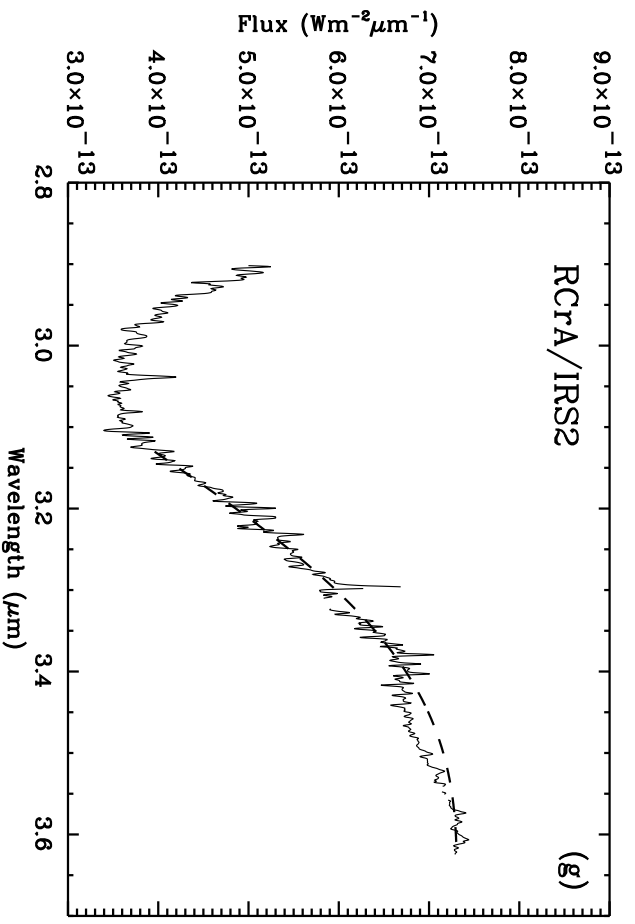
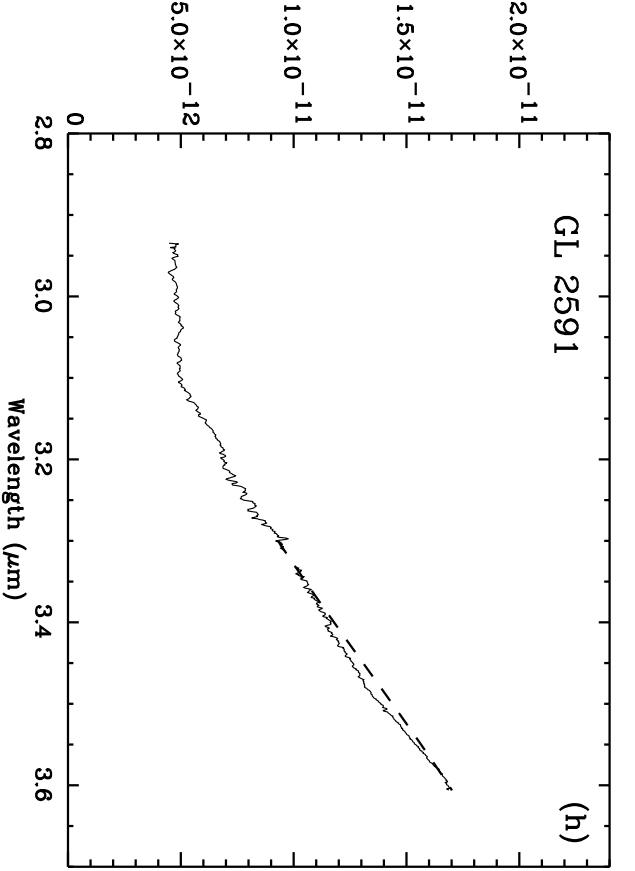
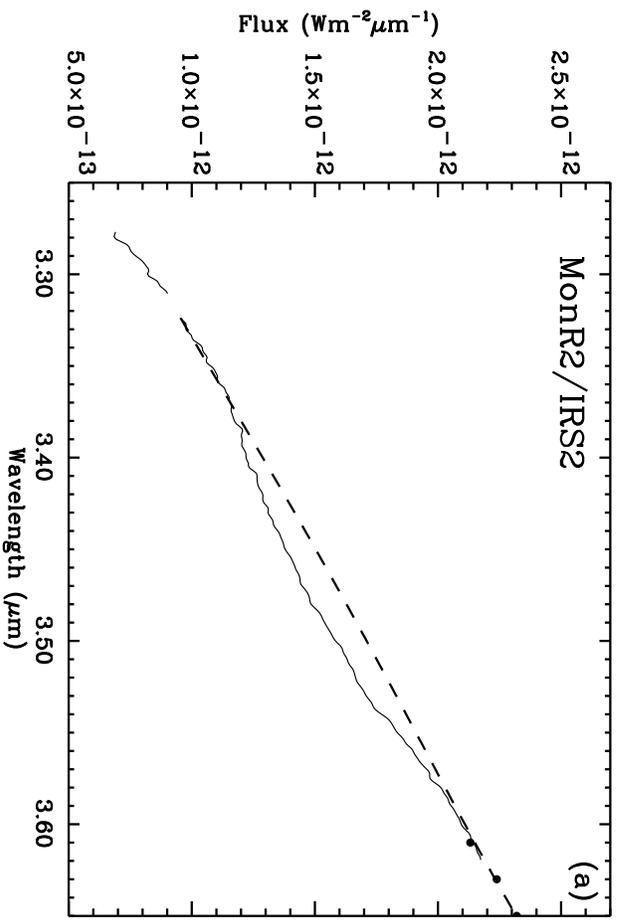
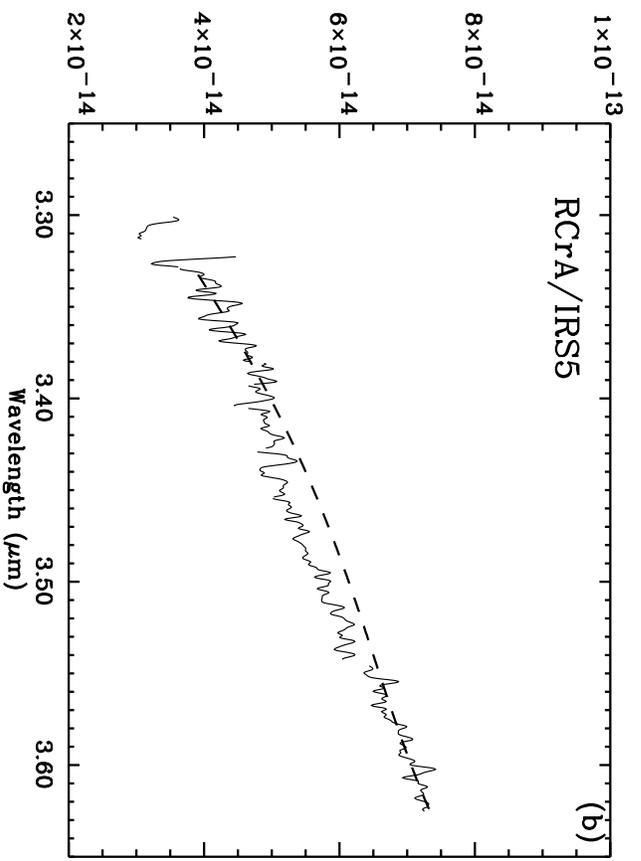
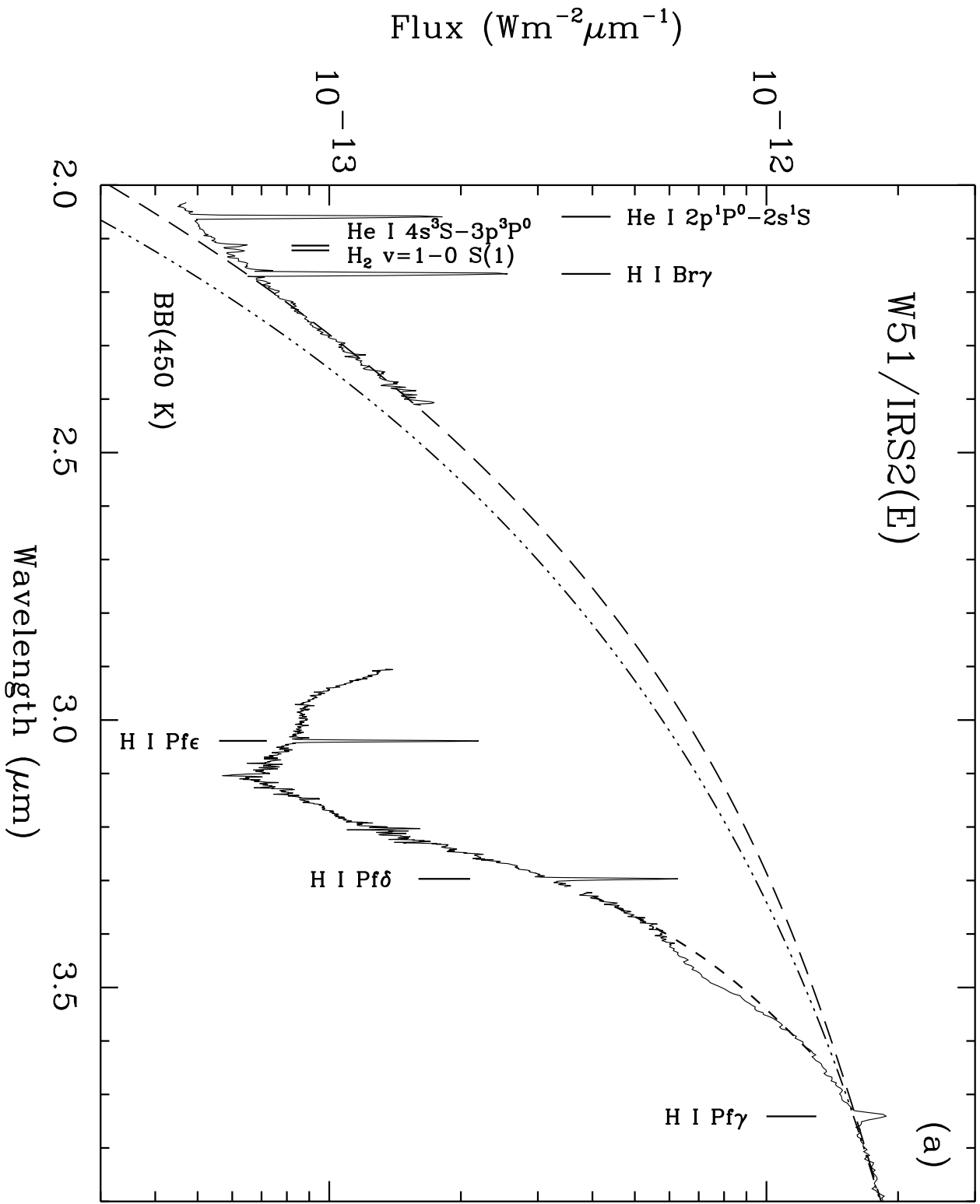
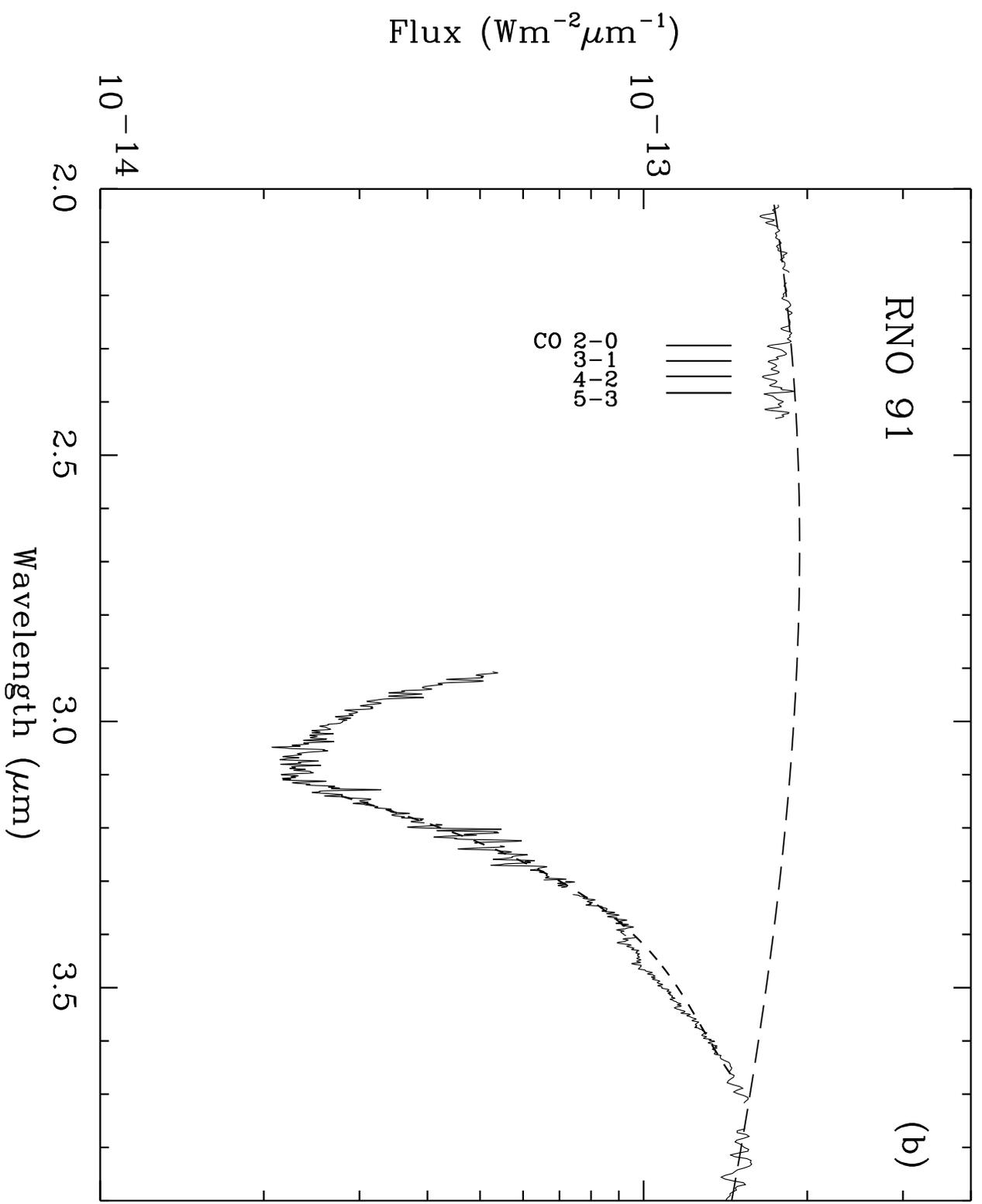
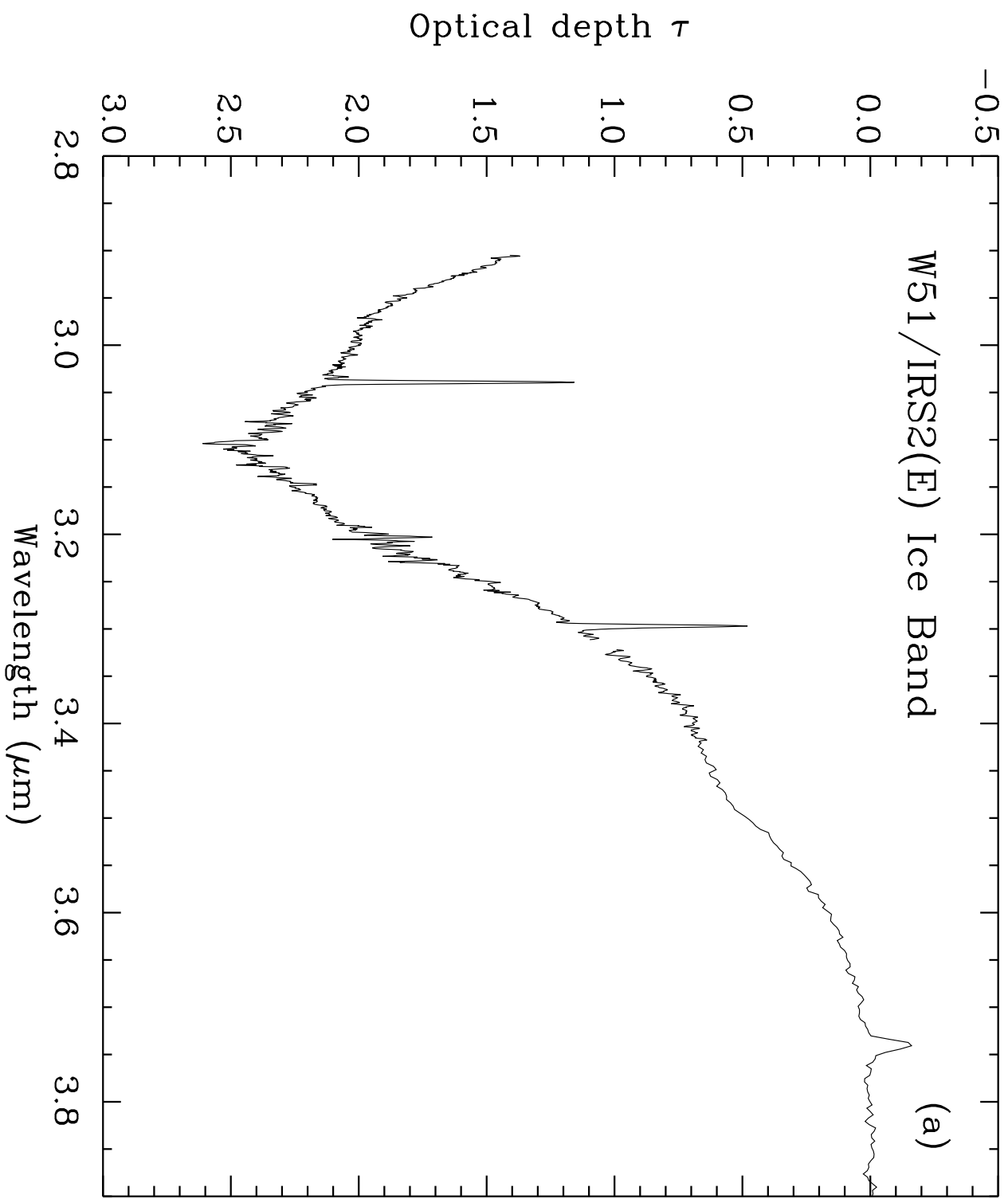
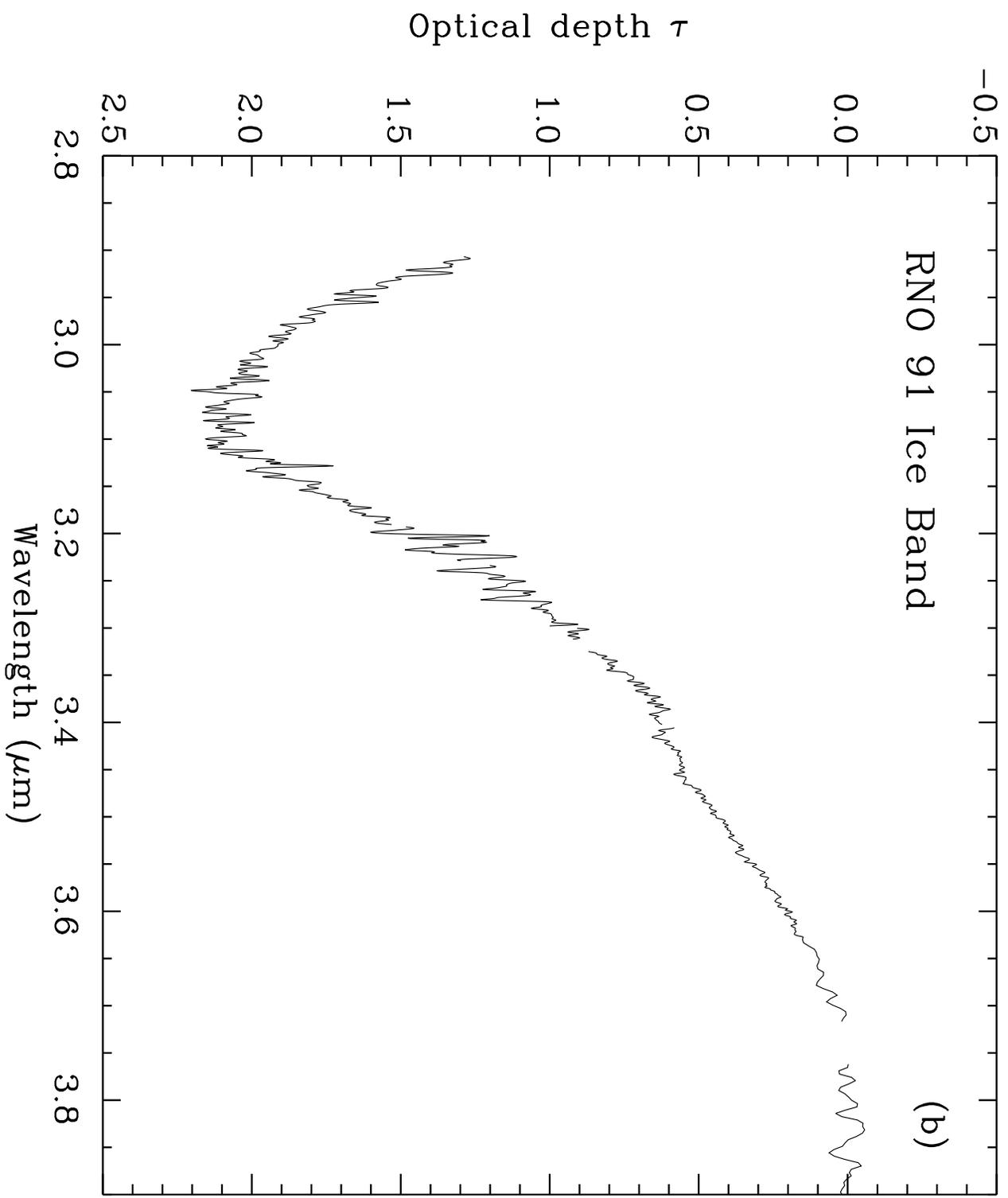
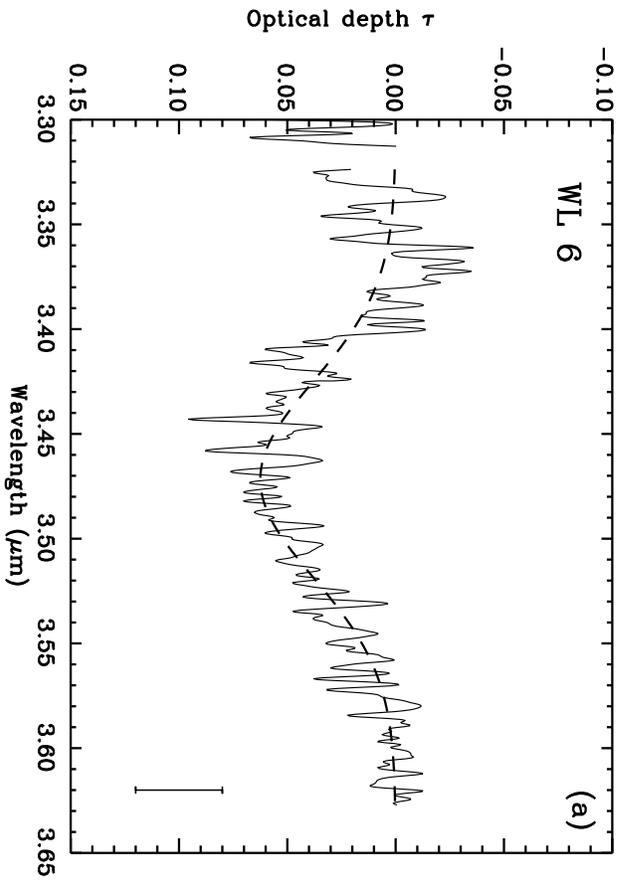
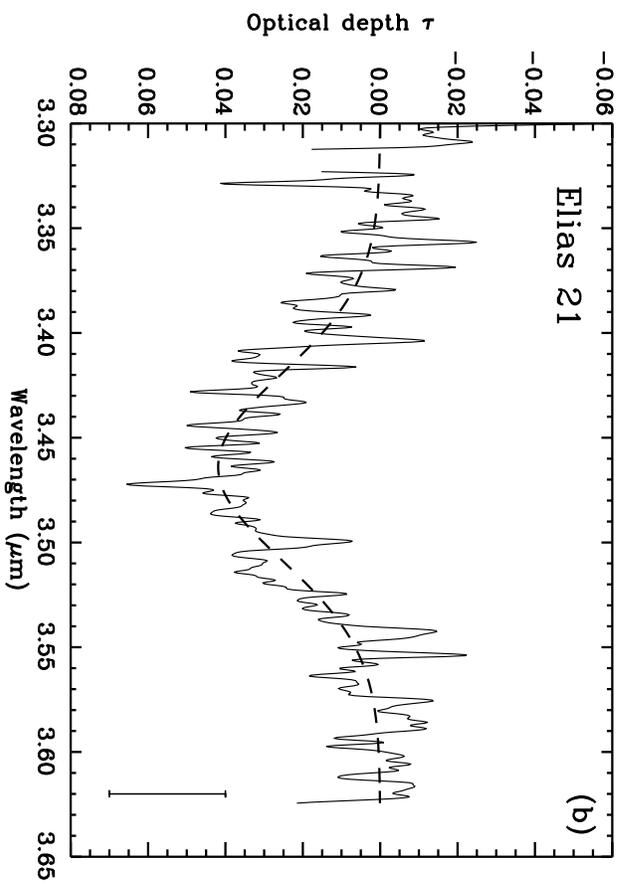
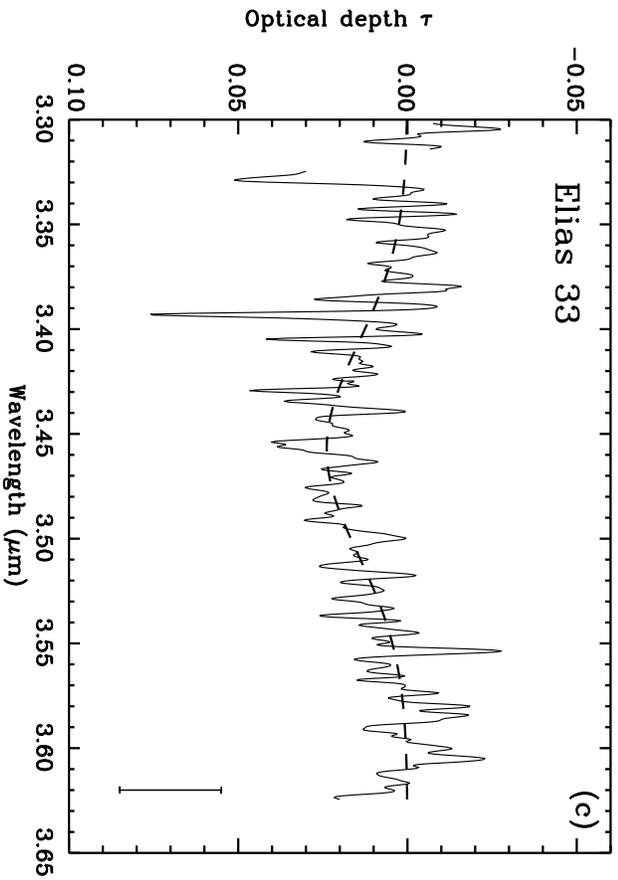
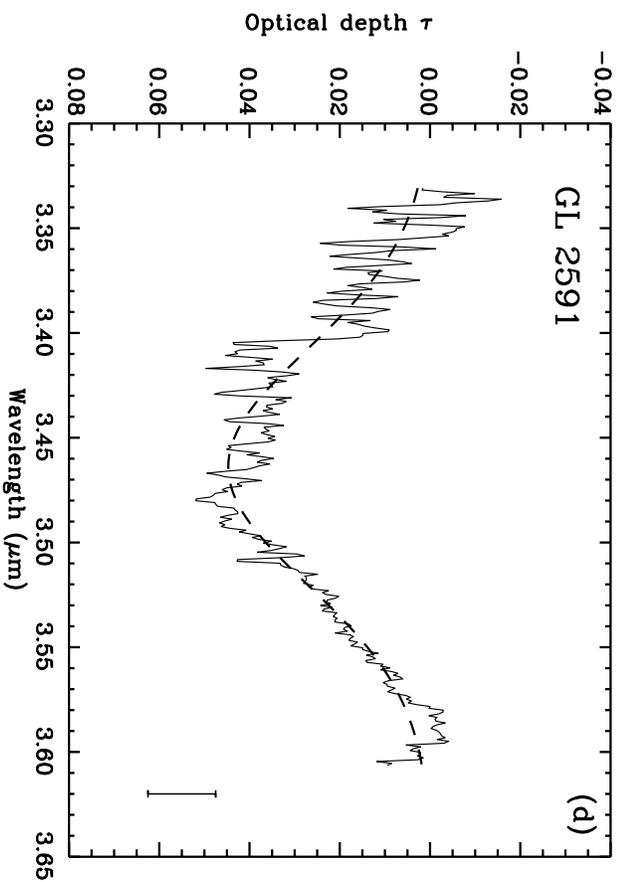
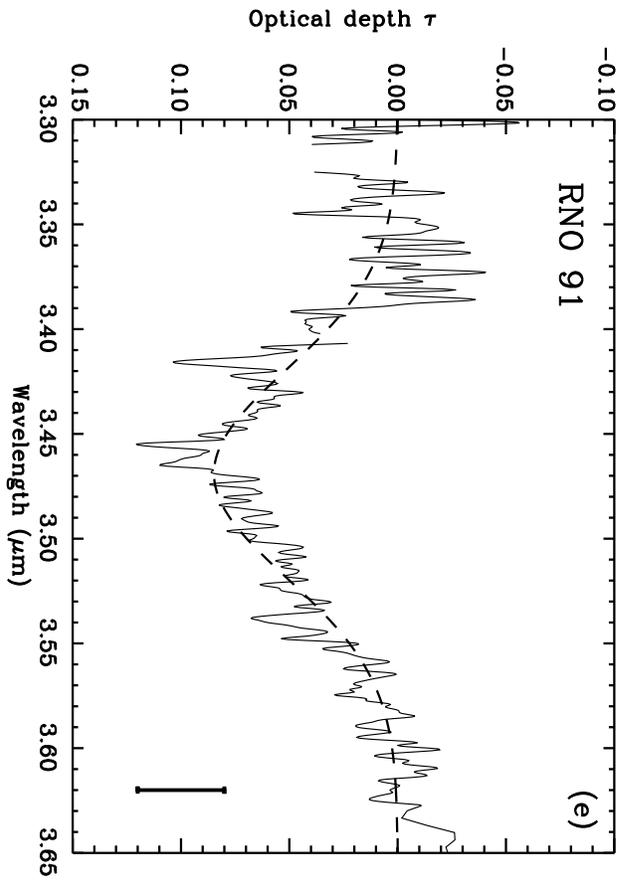
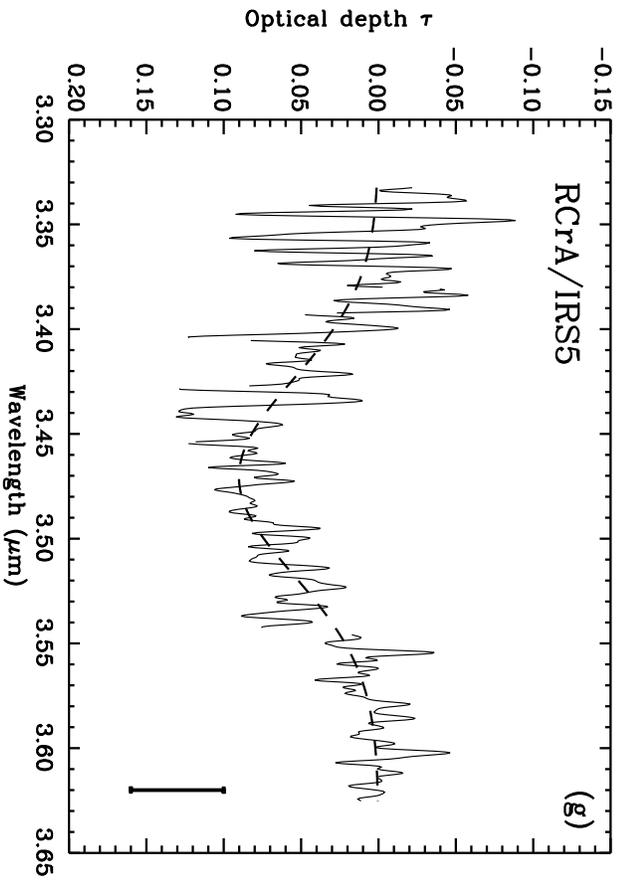
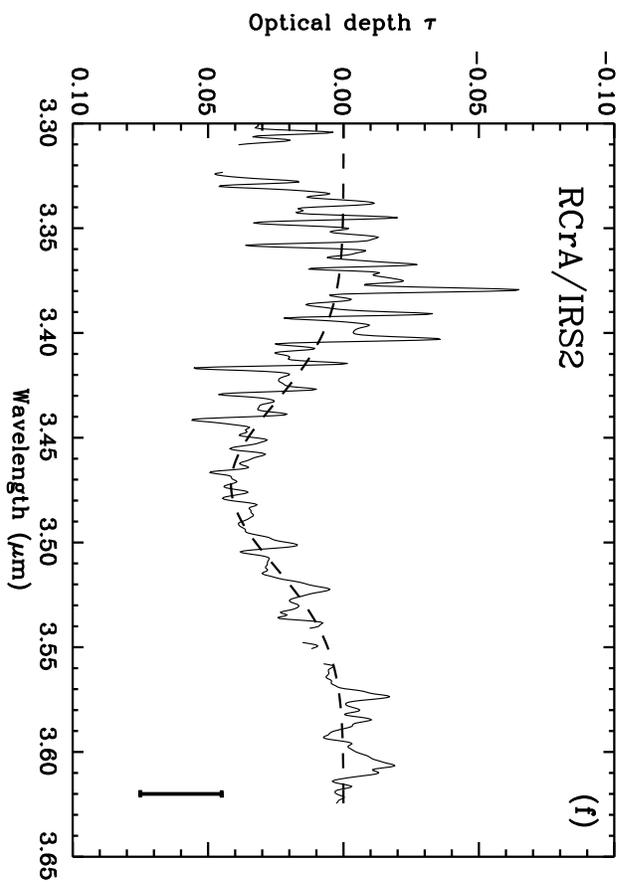
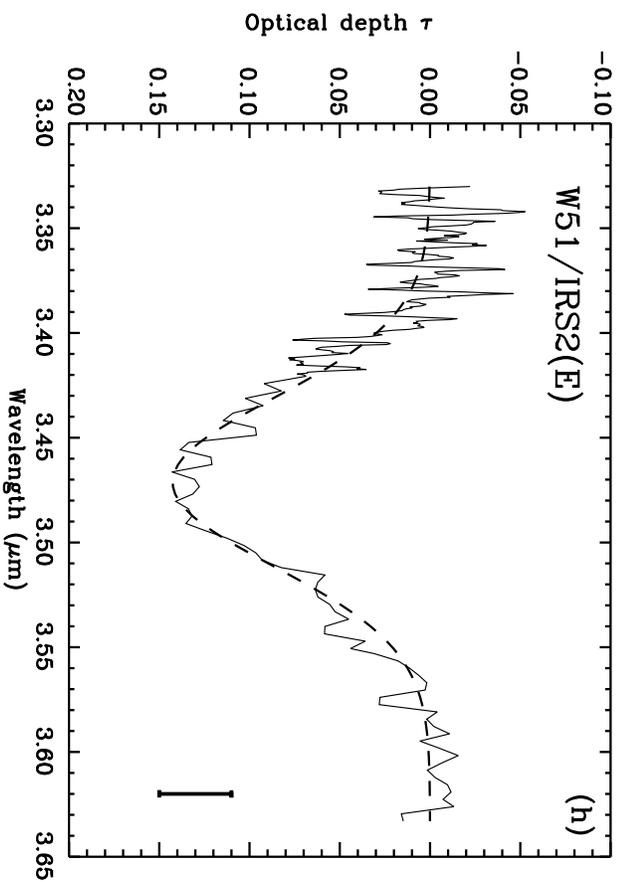
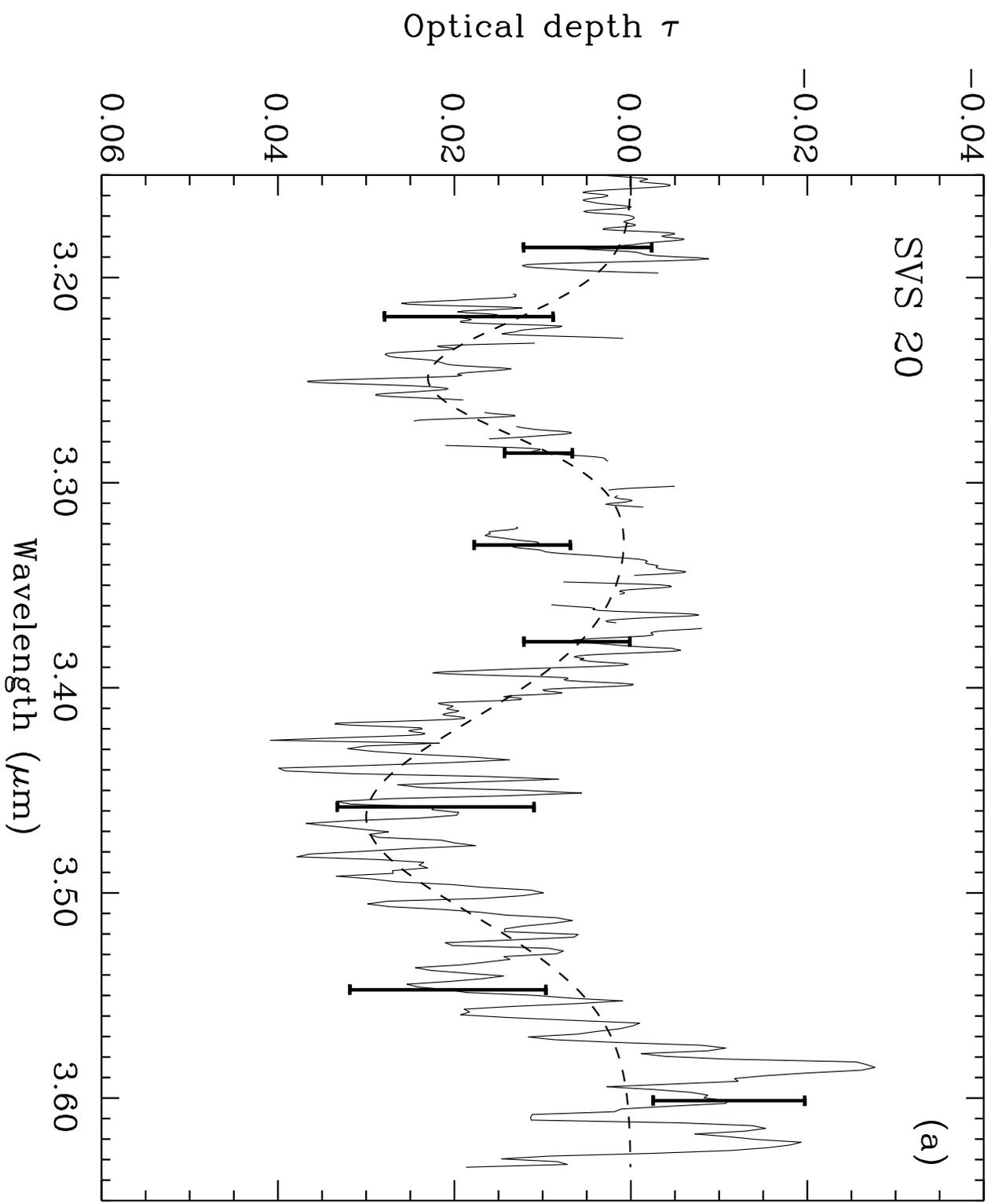
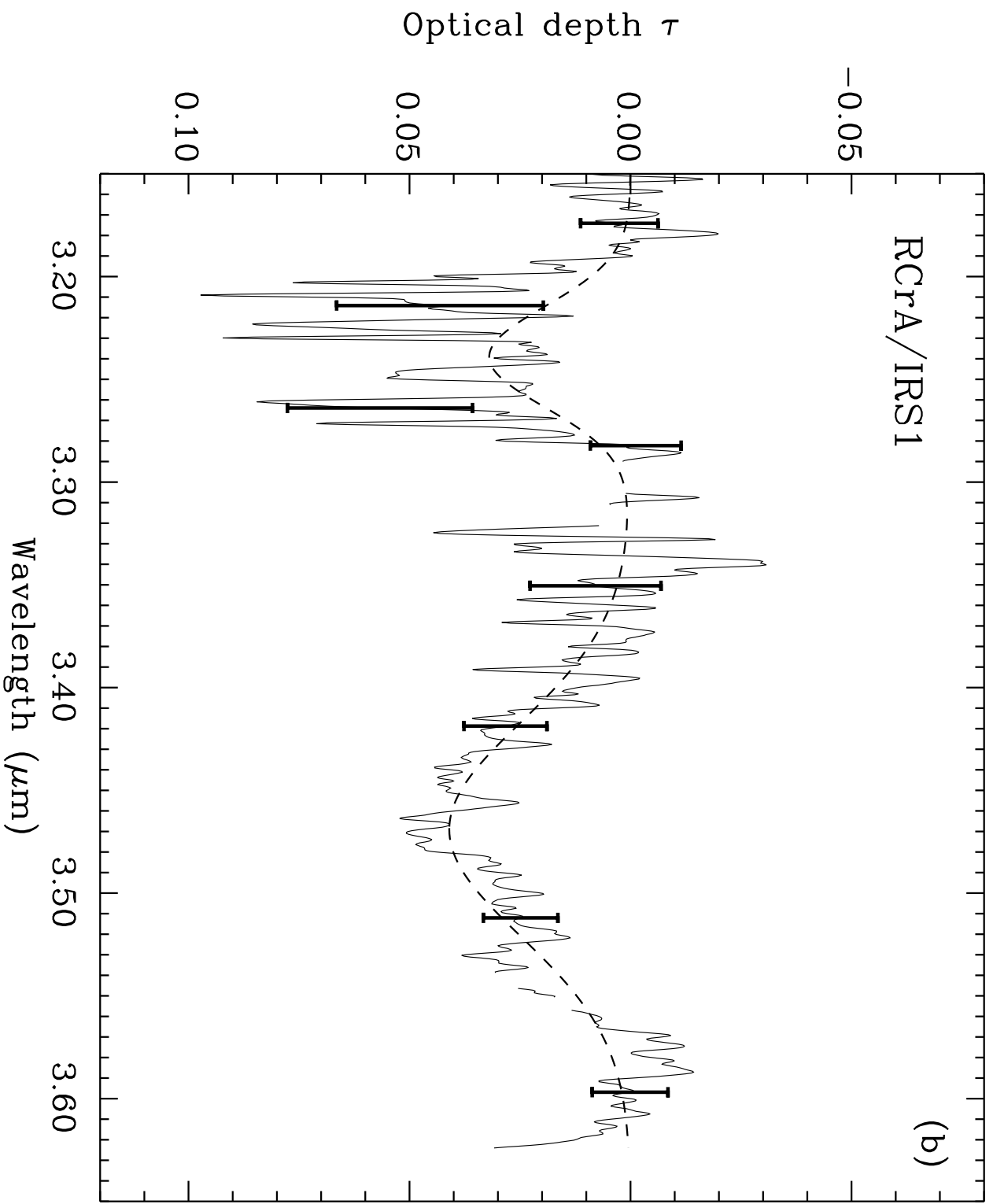
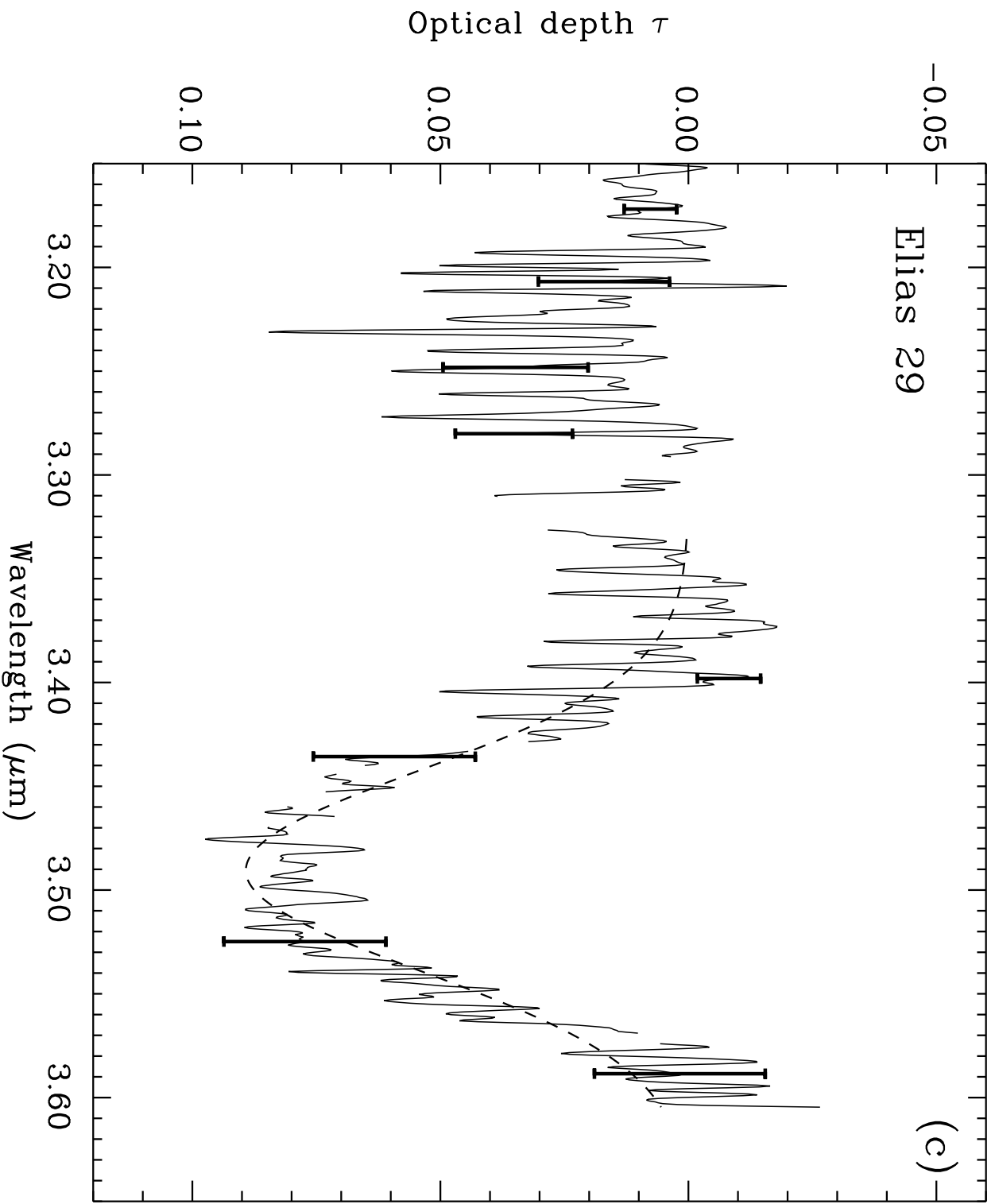
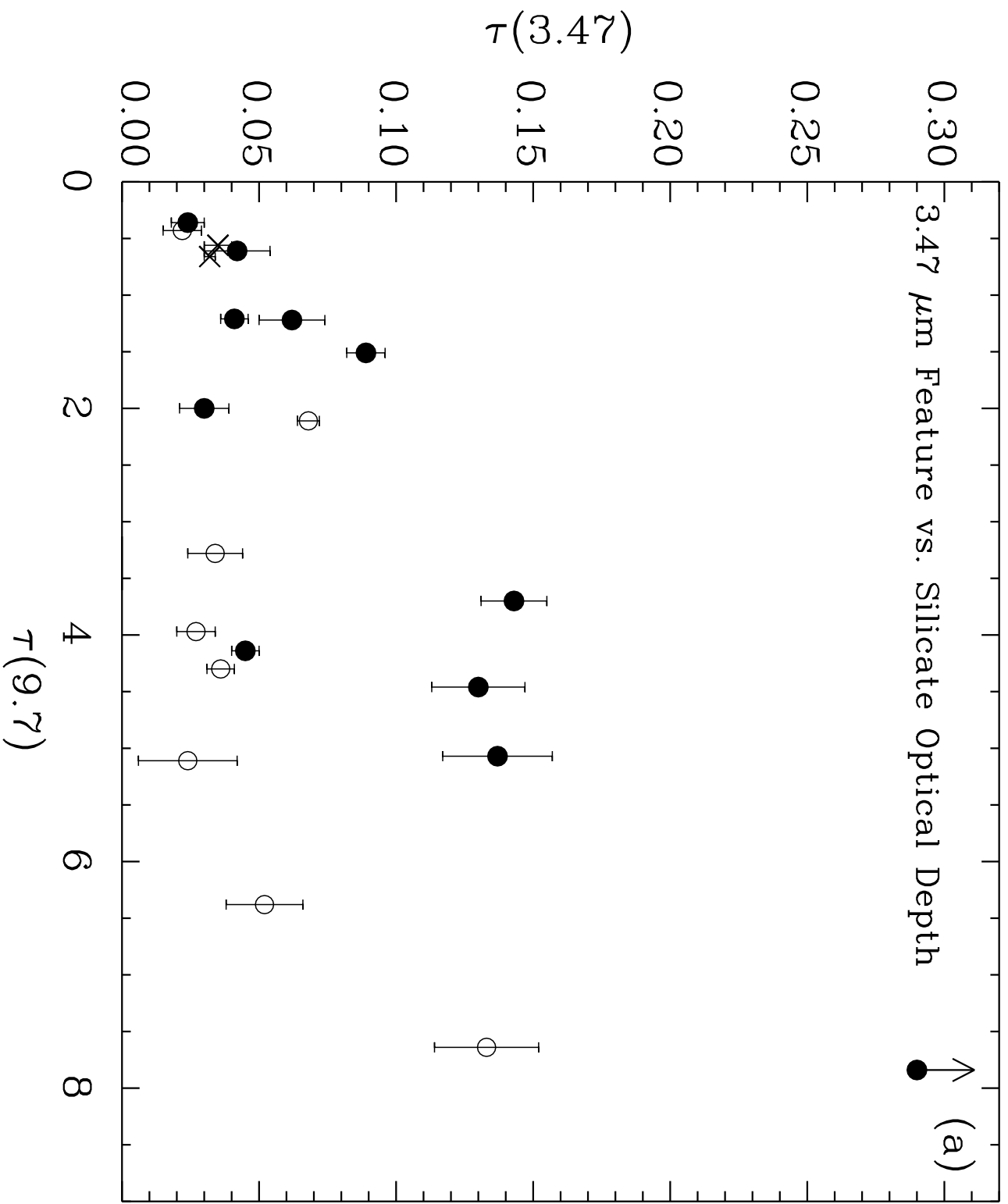
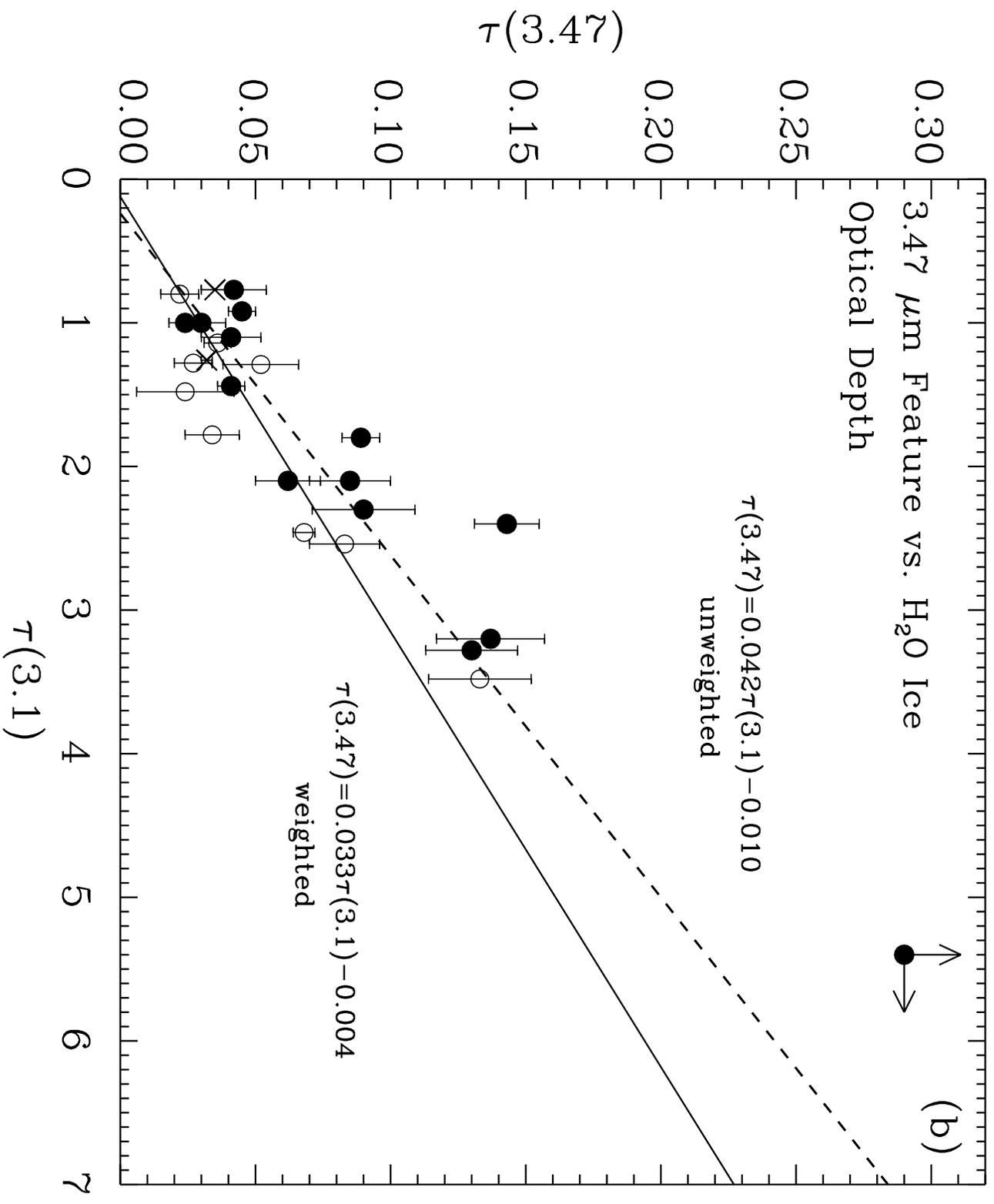
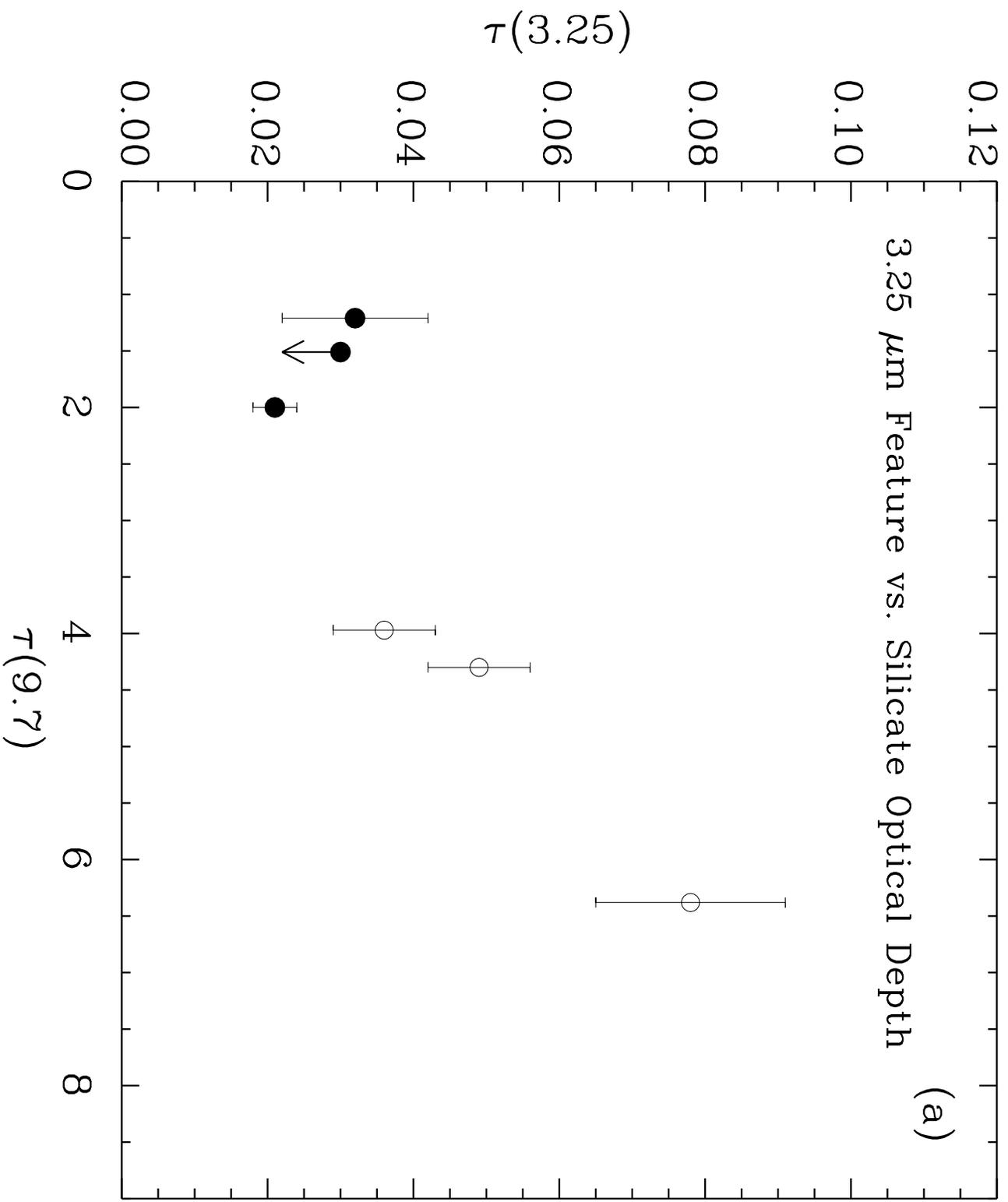
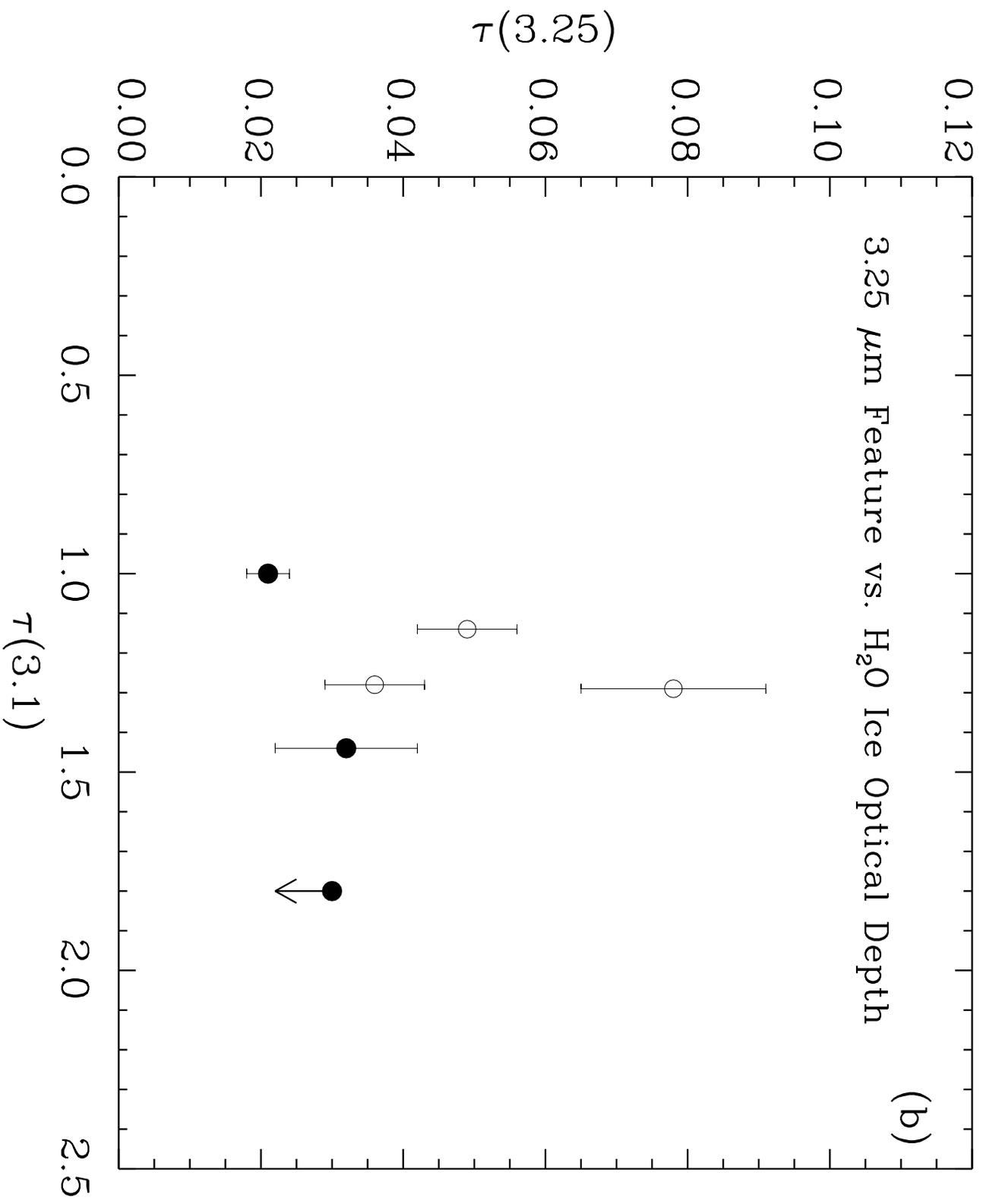
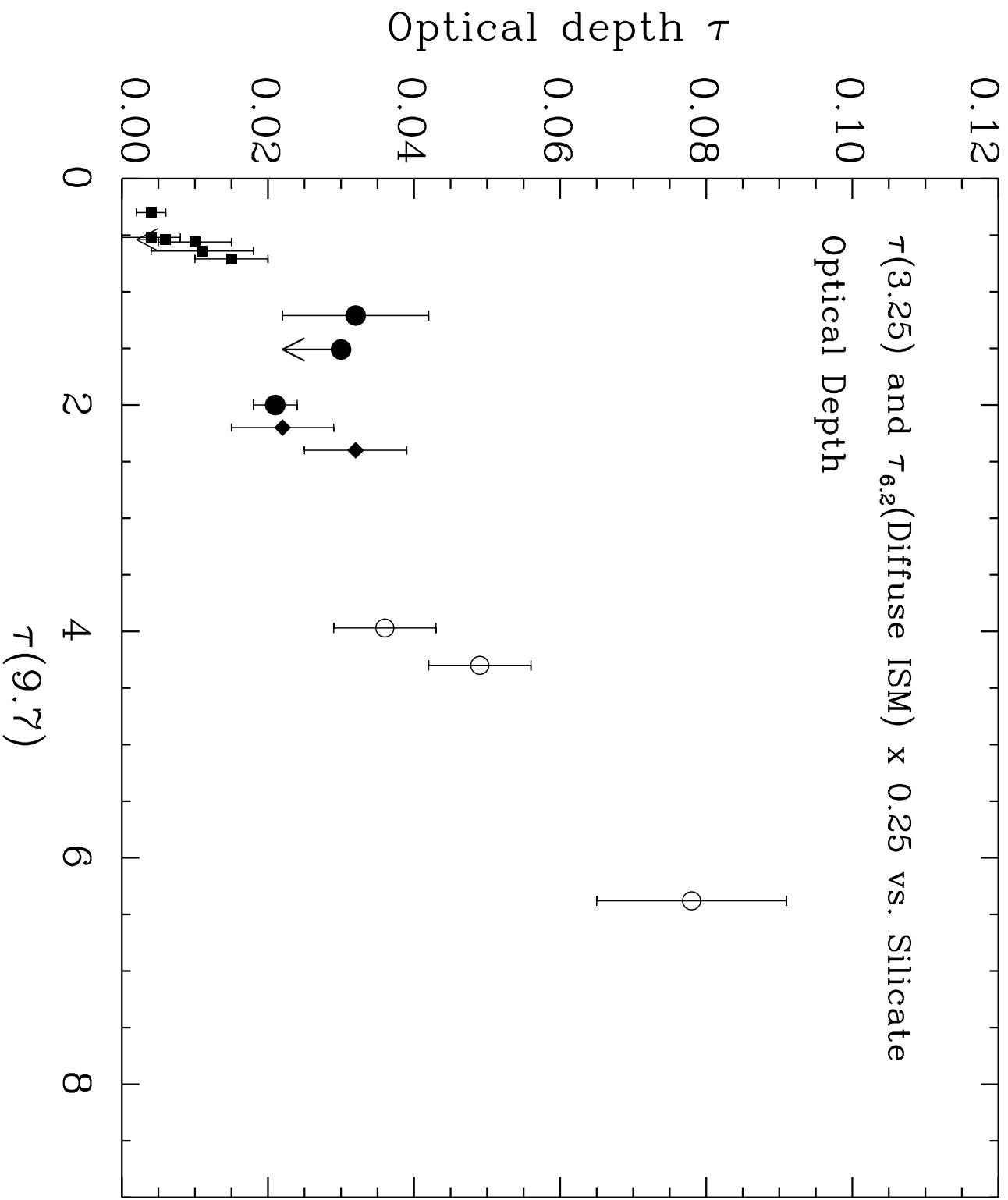
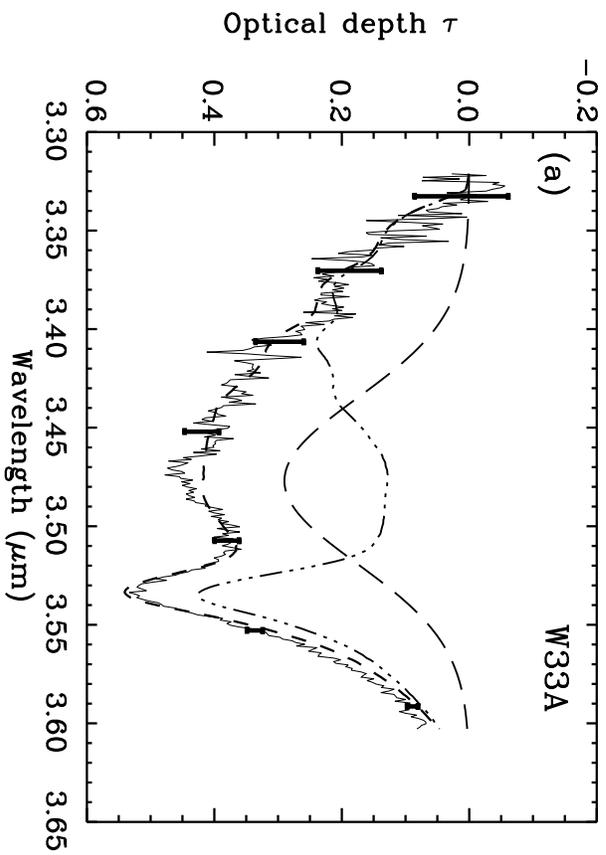
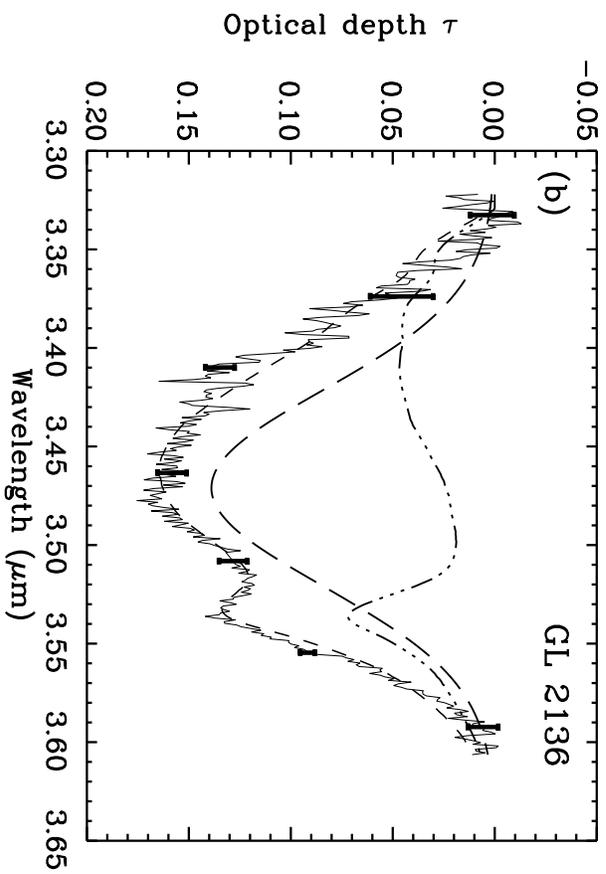
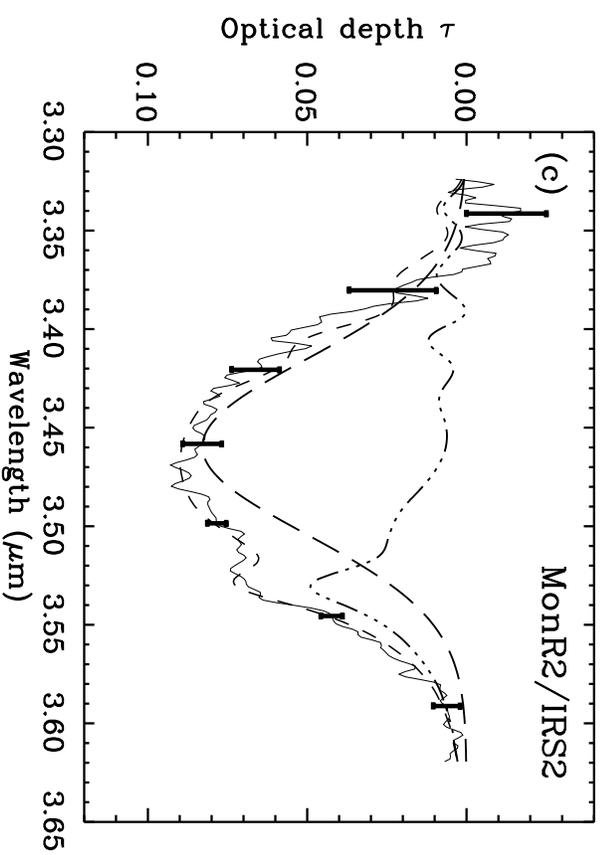
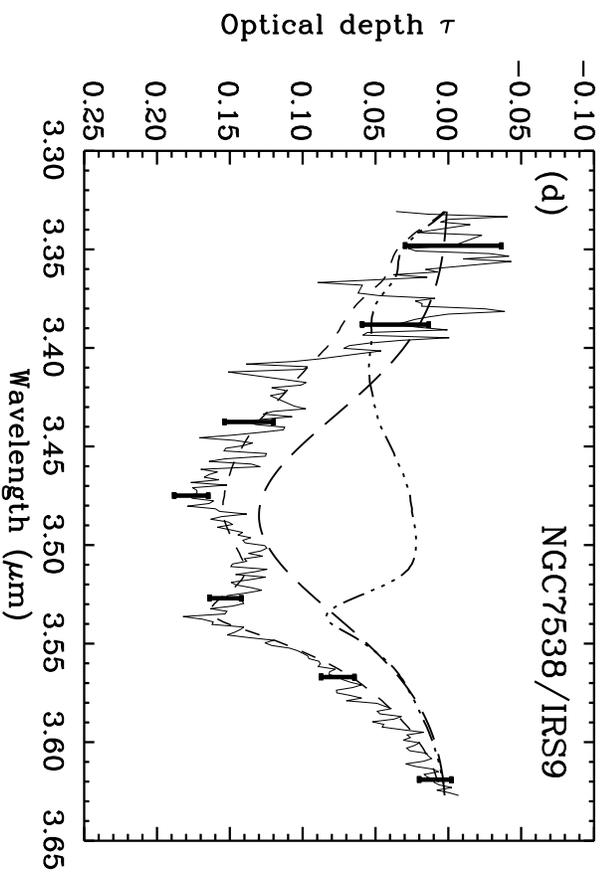
\end{document}